\documentclass{revtex4-1}
\usepackage[british]{babel}

\usepackage{graphicx}
\usepackage{epstopdf,epsfig}
\usepackage{amsmath, amsthm, amssymb,amsfonts}

\begin{document}

\title{A pressure tensor description for the time-resonant Weibel instability}

\author{{M. Sarrat}}
 \affiliation{Institut Jean Lamour, UMR 7198 CNRS - Universit\'{e} de Lorraine, France}
 \email{mathieu.sarrat@univ-lorraine.fr}
\author{D. Del Sarto}
\affiliation{Institut Jean Lamour, UMR 7198 CNRS - Universit\'{e} de Lorraine, France}%
\author{A. Ghizzo}%
\affiliation{Institut Jean Lamour, UMR 7198 CNRS - Universit\'{e} de Lorraine, France}

\date{\today}

\begin{abstract}
	We discuss a fluid model with inclusion of the complete pressure tensor dynamics for the description of Weibel type instabilities in a counterstreaming beams configuration. Differently from the case recently studied in \cite{SarratEPL}, where perturbations perpendicular to the beams were considered, here we focus only on modes propagating along the beams. Such a configuration is responsible for the growth of two kind of instabilities, the Two-Stream Instability and the Weibel instability, which in this geometry becomes ``time-resonant", \textit{i.e.} propagative. This fluid description agrees with the kinetic one and makes it possible \textit{e.g.} to identify the transition between non-propagative and propagative Weibel modes, already evidenced by \cite{LazarJPP} as a "slope-breaking" of the growth rate, in terms of a merger of two non propagative Weibel modes.
\end{abstract}

\maketitle

\section*{Introduction}

Velocity-space anisotropy-driven instabilities capable of generating strong quasi-static magnetic fields are frequent in many frameworks of plasma physics, ranging from astrophysical plasmas to laboratory laser-plasma interactions. Expressions such as ``Weibel-type" or ``Weibel-like" are frequently used as generic names for these instabilities. Examples are the pure Weibel Instability (WI) driven by a temperature anisotropy \cite{WeibelPRL} or beam-plasma instabilities like the Current Filamentation Instability (CFI), generated by a linear momentum anisotropy \cite{FriedPF}. The latter usually requires a perturbation with a wave-vector orthogonal to the beams. However, beam-plasma systems, in nature, are more generally destabilised by oblique (with respect to the beams) wave-vectors, so that classical Weibel instabilities are often in competition with the electrostatic Two-Stream Instability (TSI) or the Oblique Instability (\cite{Bret05}, \cite{Bret10}) depending on the symmetry properties of the beams and on their velocity (relativistic or not). Therefore, the family of Weibel-type instabilities includes also phenomena resulting from the combination of these two types of velocity anisotropies, such as the Weibel-CFI coupled modes or the Time-Resonant Weibel Instability (TRWI). The latter, investigated in \cite{LazarTAJ}, \cite{LazarJPP} and \cite{Ghorbanalilu}, is triggered by an excess of thermal energy perpendicularly to the direction of the electron beams and is a time-resonant, \textit{i.e.} propagative instability. At relativistic speeds, the oblique and CFI instabilities are dominant \cite{Bret10}, whereas in the non-relativistic regime the TRWI grows faster than the TSI \cite{LazarTAJ}, the oblique and the CFI.\\

In this article we focus on configurations with perturbations propagating along the beams, so that only the TRWI and the TSI can be excited. This simplification is a preliminary step to next apply the fluid model including a full pressure tensor dynamics, first investigated by \cite{BasuPOP} and by \cite{SarratEPL} for wave-vectors perpendicular to the beams, to the case of perturbations with generic oblique propagation. These modes affect the stability and the dynamics of (counterstreaming) electron beams, which in the non-relativistic regimes can be generated, \textit{e.g.}, in the nonlinear stage of three-wave parametric decays, such as Raman-type instabilities in laser-plasma interactions \cite{Masson_Laborde}. Estimating that relativistic effects become quantitatively important when electron kinetic energy is comparable to (or stronger than) their rest energy $mc^2 \sim 511$ keV (see \textit{e.g.} \cite{LazarTAJ}), a non-relativistic description of Weibel modes may be, at least qualitatively, of interest even to moderately relativistic interpenetrating intergalactic plasmas. The future extension of this model to relativistic regimes will allow broader applications to both astrophysics and high intensity laser-plasma interactions (\cite{Bret&Stockem14}, \cite{SchlickeiserTAJ}, \cite{MedvedevTAJ}).\\

A complete description of these Weibel-type instabilities requires the use of kinetic theory. Nevertheless, fluid modelisation has proven its ability to identify and understand some of their main features. \cite{FriedPF} gave a fluid picture of the pure WI, introducing the concept of CFI : the electron temperature anisotropy is replaced by a momentum anisotropy of two counterstreaming electron beams and the strong anisotropy limit of Weibel's kinetic dispersion relation is then recovered, setting up an analogy between these two driving mechanisms. This approach opened the way to the use of cold fluid models in order to study the linear phase of the relativistic and non-relativistic CFI \cite{PegoraroScripta}, possibly in presence of spatial resonances effects \cite{Califano&Pegoraro&BulanovPRE}, \cite{Califano&PrandiPRE}, and some features of its nonlinear dynamics such as the onset of secondary magnetic reconnection processes \cite{Califano&Attico}, or the coupling with the TSI in a 3D, spatially inhomogeneous configuration \cite{Califano&DelSartoPRL}.\\

These cold models are unable to reproduce important results derived from a full kinetic treatment, for example the partially electrostatic behaviour of the CFI when the beams have unequal temperatures \cite{Bret&GremilletPOPHow}, \cite{TzoufrasPRL}. Reduced kinetic models, instead, such as those based on the evolution of waterbag or multi-stream distributions \cite{InglebertPOP}, successfully describe these kinetic features. The multistream model, in particular, deals with Fried's analogy in depth by sampling the distribution function with a bunch of parallel beams, taking advantage of the potential cyclic variables of the problem (\cite{InglebertEPL}, \cite{GhizzoTrio}, \cite{GhizzoJPP}).  The fluid model we here consider, extended to include the full pressure tensor dynamics, can be compared to a reduced kinetic model. The possibility of using such a model in order to investigate Weibel instabilities was first adressed by \cite{BasuPOP}, who recovered the hydrodynamic limit of the pure WI kinetic dispersion relation. This description has been recently generalised to the thermal CFI and WI-CFI modes in a system of counterstreaming beams \cite{SarratEPL}, and it is here developed to study the onset of the TRWI. Besides resulting consistent with the kinetic description, this fluid analysis allows to characterise the transition between time-resonant and non time-resonant regimes of the instability and makes possible the identification of a second unstable branch.\\

The paper is structured as follows. In Section \ref{sec:1}, we present the model equations and the equilibrium configuration. In Section \ref{sec:2}, we perform a linear analysis for waves propagating along the beams and obtain the corresponding dispersion matrix. In Section \ref{sec:3}, we identify the hydrodynamic limit for which both the fluid and kinetic approaches are strictly equivalent. In this limit, we put in evidence the TRWI regime. In Section \ref{sec:4} we detail the analysis of the fluid dispersion relation and use it to isolate some kinds of behaviour of the TRWI and to characterise the transition between the resonant and the non-resonant regimes.

\section{The fluid model}\label{sec:1}

\subsection{Model equations}\label{sec:1.1}
We consider two non-relativistic counterstreaming electron beams ($\alpha = 1,2$) in a hydrogen plasma at time scales $\Delta t \ll \omega_{pi}^{-1}$, where $\omega_{pi}$ is the ion plasma frequency. Consequently, the ions form a neutralizing equilibrium background, with a uniform and constant density $n_i$, whose dynamics will be neglected in the following. We write the first three moments of Vlasov equation for each beam (see \textit{e.g.} \cite{DSArxiv}) :
\begin{equation}
	\frac{\partial n_{\alpha}}{\partial t} + \bf{\nabla}\cdot(n_{\alpha}\bf{u}_{\alpha}) = 0,
	\label{continuity}
\end{equation}
\begin{equation}
	\frac{\partial\boldsymbol{u}_{\alpha}}{\partial t}+\boldsymbol{u}_{\alpha}\cdot\boldsymbol{\nabla}\boldsymbol{u}_{\alpha}\\
	=-\frac{e}{m}(\boldsymbol{E}+\boldsymbol{u}_{\alpha}\times\boldsymbol{B})
	-\displaystyle{\frac{\boldsymbol{\nabla}\cdot\boldsymbol{\Pi}_{\alpha}}{m n_\alpha}},
	\label{loi_dynamique}
\end{equation}
\begin{eqnarray}
	\frac{\partial{\boldsymbol \Pi}_\alpha}{\partial t}+{\boldsymbol{\nabla}}\cdot({\boldsymbol u}_\alpha{\boldsymbol \Pi}_\alpha) 
	+ {\boldsymbol{\nabla}}{\boldsymbol u}_\alpha\cdot{\boldsymbol \Pi}_\alpha 
	+ ({\boldsymbol{\nabla}}{\boldsymbol u}_\alpha\cdot{\boldsymbol \Pi}_\alpha)^{T} \\ \mbox{}  = 
	-\frac{e}{m} ({\boldsymbol \Pi}_\alpha \times {\boldsymbol B} 
	+ ({\boldsymbol \Pi}_\alpha \times {\boldsymbol B})^{T}) - {\boldsymbol{\nabla}}\cdot{\boldsymbol Q}_\alpha,\nonumber
	\label{bilan_pression}
\end{eqnarray}
where apex ``$T$'' expresses matrix transpose. We define the pressure tensor ${\boldsymbol \Pi}_\alpha\equiv n_\alpha m(\langle {\boldsymbol v}{\boldsymbol v}\rangle_\alpha -{\boldsymbol u}_\alpha{\boldsymbol u}_\alpha)$ and the heat flux tensor 
${\boldsymbol Q}_\alpha\equiv mn_\alpha \langle ({\boldsymbol v}-{\boldsymbol u}_\alpha)({\boldsymbol v}-{\boldsymbol u}_\alpha)({\boldsymbol v}-{\boldsymbol u}_\alpha)\rangle_\alpha$. The notation $\langle...\rangle_\alpha $ indicates average in the velocity coordinate ${\boldsymbol v}$ with respect to the particle distribution function $f_\alpha({\boldsymbol x}, {\boldsymbol v})$.\\

These equations are coupled to Maxwell equations :
\begin{equation}
	\boldsymbol{\nabla}\times\boldsymbol{E}+\frac{\partial\boldsymbol{B}}{\partial t}=0, \quad 
	\boldsymbol{\nabla}\times\boldsymbol{B}=\mu_0 \boldsymbol{J}+\frac{1}{c^2}\frac{\partial\boldsymbol{E}}{\partial t},
	\label{faraday_ampère}
\end{equation}
\begin{equation}
	\boldsymbol{\nabla}\cdot\boldsymbol{B} = 0, \quad \boldsymbol{\nabla}\cdot\boldsymbol{E} = \frac{\rho}{\epsilon_0},
	\label{thomson_gauss}
\end{equation}
where $\boldsymbol{J}=-e\sum_{\alpha}n_\alpha\boldsymbol{u_\alpha}$ is the total electron current density and $\rho$ is the total charge density. The latter is related to the particle densities by $\rho = e\left(n_i - \sum_\alpha n_\alpha\right)$, where $n_i$ is the ion density and $n_\alpha$ is the $\alpha$-beam electron density.\\

Equation (\ref{bilan_pression}) requires some closure condition on the heat flux, which we choose as $\boldsymbol{\nabla}\cdot\boldsymbol{Q_\alpha} = 0$. The consistency of this choice will be shown next, by comparison with the kinetic results. On the other hand, the negligible character of $\boldsymbol{\nabla}\cdot\boldsymbol{Q_\alpha}$ in the fluid description of the pure WI has been already shown to be accurate for a wide range of wave numbers  \cite{SarratEPL}.

\subsection{Equilibrium configuration}\label{sec:1.2}

\begin{figure}
\centerline{\includegraphics[width=8.cm]{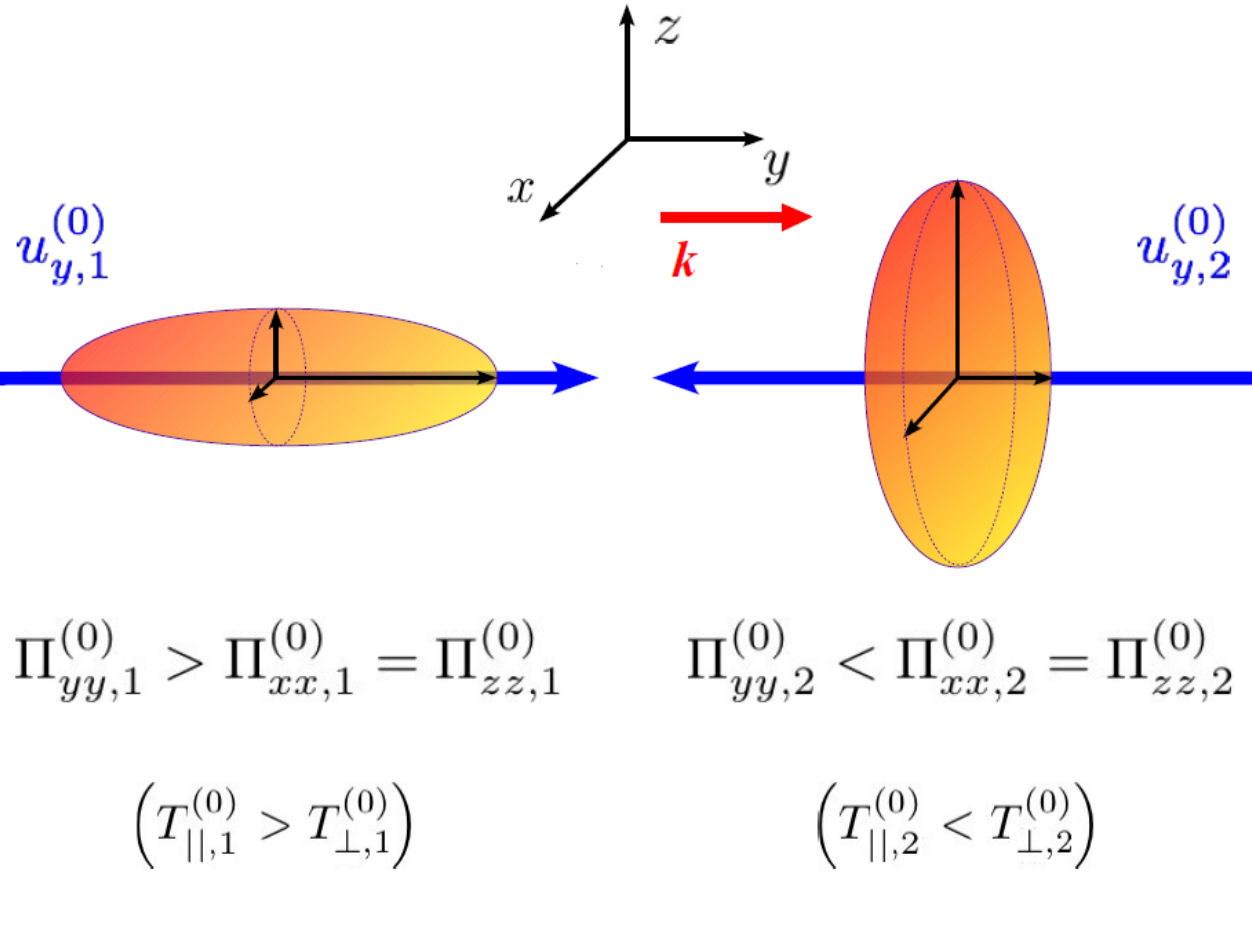}}
\caption{Two asymmetrical and counterstreaming beams perturbed by a wave propagating along them. Each beam presents an initial pressure anisotropy between parallel ($\parallel$) and perpendicular ($\perp$) - to the wave-vector - components. In the limit of vanishing initial velocities, $u_{y,1}^{(0)}=u_{y,2}^{(0)}=0$, as the pure WI requires an excess of perpendicular thermal energy, the population $(1)$ would be stable to a pure Weibel Instability whereas the population $(2)$ would be unstable.}
\label{fig:1}
\end{figure}

At equilibrium, we suppose the beams along the $y$-axis (see Figure \ref{fig:1}) and represented by their own set of fluid variables : density $n_\alpha^{(0)}$, fluid velocity $\bf{u}_\alpha^{(0)} = u_{y,\alpha}^{(0)}\bf{e}_y$ and pressure tensor $\bf{\Pi}^{(0)}_{\alpha}$. These quantities are potentially different for each beam. Equilibrium velocities and density are constrained in order to avoid equilibrium electromagnetic fields : the quasi-neutrality is ensured by $\sum_\alpha n_\alpha^{(0)}=n_i=n^{(0)}$ and the total electron current is set to zero, that is $\sum_{\alpha}n_{\alpha}^{(0)}u_{\alpha,y}^{(0)} = 0$. We define the squared plasma frequencies for each beam at equilibrium and for the complete electron system :
\begin{equation}
\omega^2_{pe,\alpha} \equiv \frac{n_\alpha^{(0)}e^2}{m\epsilon_0},\qquad \omega_{pe}^2\equiv
\omega_{pe,1}^2+\omega_{pe,2}^2 = \frac{c^2}{d_e^2},
\label{eq:plasma}
\end{equation}
with $d_e$ the electron skin depth.\\

We consider an equilibrium pressure tensor uniform in space but whose initial diagonal components are in principle different one from each other, 
\begin{equation}
\setlength{\arraycolsep}{5pt}
\renewcommand{\arraystretch}{1.5}
\boldsymbol{\Pi}^{(0)}_{\alpha} = \left[
\begin{array}{ccc}
  	\Pi_{xx,\alpha}^{(0)}  	&  0               			&  0  						\\
  	0               			&  \Pi_{yy,\alpha}^{(0)}  	&  0  						\\
  	0               			&  0               			&  \Pi_{zz,\alpha}^{(0)}  	\\
\end{array}  \right],
\label{eq:pressure_eq}
\end{equation}
and we define the squared velocities,
\begin{equation}
	c^2_{x,\alpha} \equiv \frac{\Pi_{xx,\alpha}^{(0)}}{m n_\alpha^{(0)}},\qquad
	c^2_{y,\alpha} \equiv \frac{\Pi_{yy,\alpha}^{(0)}}{m n_\alpha^{(0)}},\qquad
	c^2_{z,\alpha} \equiv \frac{\Pi_{zz,\alpha}^{(0)}}{m n_\alpha^{(0)}}.
\label{eq:v_sound}
\end{equation}

Similar anisotopic pressure configurations, though generally non-uniform in space, are evidenced by both direct satellite measurements and kinetic simulations of the solar-wind turbulence (see for example \cite{servidio2012local}, \cite{servidio2015kinetic}, \cite{Franci}) or magnetic reconnection (see for example \cite{scudder2008illuminating}, \cite{scudder2012first}, \cite{scudder2016collisionless}). Even if the mechanism of temperature anisotropisation in each context is still matter of investigation, the dynamical action of the fluid strain on the pressure tensor components, recently identified  as a possible \textit{ab initio} source of non-gyrotropic anisotropy, seems to be a good candidate to explain temperature anisotropies measured \textit{e.g.} in kinetic turbulence (\cite{DS_PRE}, \cite{Franci}). Hereafter we assume the configuration of Eq.(\ref{eq:pressure_eq}) as a general equilibrium condition potentially unstable to Weibel-type modes, which can be related to a three-Maxwellian particle distribution with kinetic temperatures $k_{_B}T_{i,\alpha}\equiv \Pi_{ii,\alpha}^{(0)}/n_\alpha^{(0)}$ for $i = x,y,z$,
\begin{equation}
	f^{(0)}= n_0 \prod_{i = x}^z\left(\frac{m}{2\pi k_{_B} T_i}\right)^{\frac{1}{2}}\text{exp}\left(-\frac{mv_i^2}{2 k_{_B} T_i}\right) = n_0 \prod_{i = x}^z \frac{1}{\sqrt{2\pi}c_{i,\alpha}}\text{exp}\left(-\frac{v_i^2}{2c_{i,\alpha}^2} \right).
\label{eq:fMaxwell3D}
\end{equation}

This initial configuration is unstable because of two kinds of anisotropy : one in pressure ($\Pi_{xx,\alpha}^{(0)} \neq \Pi_{yy,\alpha}^{(0)} \neq \Pi_{zz,\alpha}^{(0)} $), the other in momentum (\textit{i.e.} the counterstreaming beams configuration).\\
The first one triggers the pure WI when the thermal spread transverse to a wave vector $\boldsymbol{k}$ is greater than the one in the parallel direction.   
The second one generates instability whatever the wave vector orientation, due to the natural repulsion between two counterstreaming electric currents. A perturbation orthogonal to the flow triggers the CFI whereas a perturbation along the flow is responsible for the TSI. A slanting perturbation gives birth to the so-called oblique modes. The CFI configuration has been widely examinated, in particular the coupling between the CFI and the pure WI modes. This coupling was recently investigated from a kinetic point of view by \cite{LazarTAJ} in the non-relativistic context, by \cite{Bret10} in the relativistic one, and using the non-relativistic fluid pressure tensor description by \cite{SarratEPL}.\\

In this work, we consider perturbations with wave-vector $\boldsymbol{k} = k_y\boldsymbol{e}_y$, propagating along the beams in presence of an initial pressure anisotropy (see Figure \ref{fig:1}). This configuration gives birth to temperature anisotropy driven modes, not coupled with the TSI and sometimes propagative (\textit{i.e.} the TRWI, \cite{LazarJPP}).

\section{Dispersion matrix}\label{sec:2}
We perturb the  equilibrium with Fourier modes $\propto \text{exp}\left[-i(\omega t - k_y y) \right]$, where we define the complex frequency $\omega \equiv \omega_r + i\gamma$, consisting of a real, propagative part, $\omega_r$,  and of a growth (or damping) rate, $\gamma$.

\subsection{Linearisation}\label{sec:2.1}

Starting with the continuity equation (\ref{continuity}), we write
\begin{equation}
	n_\alpha^{(1)} = n_\alpha^{(0)} \frac{k_y u_{y,\alpha}^{(1)}}{w_\alpha},
	\label{eq:continuity_lin}
\end{equation}	
	where we introduce the notation 
\begin{equation}
	w_\alpha \equiv\omega - \boldsymbol{k}\cdot\boldsymbol{u}_\alpha^{(0)} = \omega - k_y u_{y,\alpha}^{(0)},
\end{equation}
In a vector form, the linear momentum balance writes
\begin{equation}
	\boldsymbol{u}_\alpha^{(1)} = -\frac{ie}{m w_\alpha}
		\left(\boldsymbol{E}^{(1)}+\boldsymbol{u}_\alpha^{(0)}\times\boldsymbol{B}^{(1)}\right)
		+ \frac{\boldsymbol{k}}{w_\alpha}\cdot\frac{\boldsymbol{\Pi}^{(1)}_{\alpha}}{mn_\alpha^{(0)}},
	\label{eq:euler_lin}
\end{equation}
with the following expressions for the diagonal components of $\boldsymbol{\Pi}^{(1)}_{\alpha}$ (cf. Eq. (\ref{bilan_pression})),
\begin{equation}
	\Pi_{xx,\alpha}^{(1)} = \frac{k_y}{w_\alpha}\Pi_{xx,\alpha}^{(0)}u_{y,\alpha}^{(1)},\quad
	\Pi_{zz,\alpha}^{(1)} = \frac{k_y}{w_\alpha}\Pi_{zz,\alpha}^{(0)}u_{y,\alpha}^{(1)},
	\label{eq:pixxzz_lin}
\end{equation}
\begin{equation}
	\Pi_{yy,\alpha}^{(1)} = 3\frac{k_y}{w_\alpha}\Pi_{yy,\alpha}^{(0)}u_{y,\alpha}^{(1)},
	\label{eq:piyy_lin}
\end{equation}
and for the non diagonal components,
\begin{equation}
	\Pi_{xz,\alpha}^{(1)} = 0,
	\label{eq:pixz_lin}
\end{equation}
\begin{equation}
	\Pi_{xy,\alpha}^{(1)} = \frac{k_y}{w_\alpha}\Pi_{yy,\alpha}^{(0)}u_{x,\alpha}^{(1)}
	- \frac{ie}{m}\frac{\Pi_{yy,\alpha}^{(0)}-\Pi_{xx,\alpha}^{(0)}}{w_\alpha}B_z^{(1)},
	\label{eq:pixy_lin}
\end{equation}
\begin{equation}
	\Pi_{yz,\alpha}^{(1)} = \frac{k_y}{w_\alpha}\Pi_{yy,\alpha}^{(0)}u_{z,\alpha}^{(1)}
	- \frac{ie}{m}\frac{\Pi_{zz,\alpha}^{(0)}-\Pi_{yy,\alpha}^{(0)}}{w_\alpha}B_x^{(1)}.\\
	\label{eq:piyz_lin}
\end{equation}
The Maxwell-Faraday equation allows us to eliminate the magnetic field : $B_x^{(1)}=k_y E_z^{(1)}/\omega$ and $B_z^{(1)}=-k_y E_x^{(1)}/\omega$.\\

The evolution of $u_{y,\alpha}^{(1)}$ completely determines that of $\Pi_{xx,\alpha}^{(1)}$ and $\Pi_{zz,\alpha}^{(1)}$ via plasma compressibility effects, but these $\boldsymbol{\Pi_\alpha}$ components are not linearly involved in the fluid velocity dynamics. Notice that the $\Pi_{xz,\alpha}$ component remains zero during the whole linear evolution and that the two other non-diagonal components can evolve even if the equilibrium pressure is isotropic due to the first r.h.s term in equations \eqref{eq:pixy_lin} and \eqref{eq:piyz_lin}. The role of an initial pressure anisotropy on the magnetic field generation clearly appears in the second r.h.s. of the two latter equations.\\

\subsection{Dispersion relations}\label{sec:2.2}

Combining linearised Maxwell and fluid equations leads to the generalised dispersion relation, $[\bf{D}]\cdot\bf{E}^{(1)}=0$, with $[\boldsymbol{D}]$ the dispersion matrix,
\begin{equation}
	\setlength{\arraycolsep}{5pt}
	\renewcommand{\arraystretch}{1.5}
	\boldsymbol{D} = \left[
	\begin{array}{ccc}
  	D_{xx}  	&  0 		&  0  		\\
  	0	  	&  D_{yy} 	&  0  		\\
  	0     	&  0       	&  D_{zz}	\\
	\end{array}  \right],
	\label{eq:generalized_Dmatrix}
\end{equation}
whose elements $D_{ij}$ are given by
\begin{equation}
	D_{xx} = \frac{\omega^2}{c^2}-k_y^2 - \sum_\alpha \frac{\omega^2_{pe,\alpha}}{c^2}
	\frac{w_\alpha^2+k_y^2(c^2_{x,\alpha}-c^2_{y,\alpha})}{w_\alpha^2-k_y^2c^2_{y,\alpha}},
	\label{eq:Dxx_tsiconf}
\end{equation}
\begin{equation}
	D_{yy} = \frac{\omega^2}{c^2} - \frac{\omega^2}{c^2}\sum_\alpha\frac{\omega^2_{pe,\alpha}}{w_\alpha^2-3k_y^2c^2_{y,\alpha}},
	\label{eq:Dyy_tsiconf}
\end{equation}
and
\begin{equation}
	D_{zz} = \frac{\omega^2}{c^2}-k_y^2 - \sum_\alpha \frac{\omega^2_{pe,\alpha}}{c^2}
	\frac{w_\alpha^2+k_y^2(c^2_{z,\alpha}-c^2_{y,\alpha})}{w_\alpha^2-k_y^2c^2_{y,\alpha}}.
	\label{eq:Dzz_tsiconf}
\end{equation}

Like in the kinetic framework \cite{LazarTAJ}, we obtain three decoupled dispersion relations in the linear regime, given by $\mbox{Det}[\boldsymbol{D}]=0$. The relation $D_{yy}=0$ corresponds to plasma oscillations modified by the existence of the beams and damped by thermal effects : this is the fluid TSI dispersion relation. Notice that $D_{yy}$ does not depend on the pressure anisotropy, but only on the thermal spread along the beams direction, like in the kinetic framework. The dispersion relations given by $D_{xx} = 0$ and $D_{zz} = 0$ are identical after replacing $c^2_{x,\alpha}$ by $c^2_{z,\alpha}$ and vice-versa. The dispersion relation $D_{xx} = 0$ depends on the pressure anisotropy and describes the time-resonant Weibel-type modes we focus on in the following.\\

In order to discuss the transition between the non-propagative and the time-resonant character of the Weibel dispersion relation, we make the further simplifying assumption to consider the special case of two symmetrical beams. We then write $n_1^{(0)} = n_2^{(0)} = n^{(0)}/2$, $u_{y,2}^{(0)}=-u_{y,1}^{(0)}=-u_0$, $c_{z,\alpha} = c_z$, $c_{y,\alpha} = c_y$, $c_{x,\alpha} = c_x \; \forall \alpha$ and $c_y \neq c_x \neq c_z$. The equations (\ref{eq:Dxx_tsiconf}) and (\ref{eq:Dyy_tsiconf}) respectively become 
\begin{equation}
	\frac{\omega^2}{c^2}-k_y^2 - \frac{\omega^2_{pe}}{2c^2}\sum_\alpha 
	\frac{\displaystyle{1+\frac{k_y^2(c^2_{x}-c^2_{y})}{w_\alpha^2}}}{\displaystyle{1-\frac{k_y^2c^2_{y}}{w_\alpha^2}}}=0
	\label{eq:tsiconf_Dxx_DLposition}
\end{equation}
and
\begin{equation}
	1 - \frac{\omega^2_{pe}}{2}\sum_\alpha\frac{1}{w_\alpha^2}\frac{1}{\displaystyle{1-\frac{3k_y^2c^2_{y}}{w_\alpha^2}}}  = 0.
	\label{eq:tsiconf_Dyy_DLposition}
\end{equation}

\section{The Hydrodynamic Limit}\label{sec:3}

We introduce the parameter 
\begin{equation}
	\epsilon_\alpha \equiv \frac{k_yc_{y,\alpha}}{w_{\alpha}},
\end{equation}
which satisfies the criterion
\begin{equation}
	|\epsilon_{\alpha}| \ll 1
	\label{eq:criterion}
\end{equation}
at the hydrodynamic limit.
In the case of perpendicular propagation of WI-CFI coupled modes, it becomes the condition $k_yc_{y,\alpha}/|\omega| \ll 1$. Once satisfied for each beam, this latter criterion has been shown to grant an identical description of the linear WI-CFI modes in the kinetic and in the extended fluid models \cite{SarratEPL}.

%As discussed in \citet{SarratEPL}, we introduce the hydrodynamic limit as satisfying the criterion
%\begin{equation}
%	\epsilon_\alpha \equiv \frac{k_yc_{y,\alpha}}{|w_{\alpha}|} = \frac{k_yc_{y,\alpha}}{\sqrt{(\omega_r - k_yu_{0,\alpha})^2+\gamma^2}} \ll 1.
%	\label{eq:criterion}
%\end{equation}
%In the case of perpendicular propagation of WI-CFI coupled modes, it becomes the condition $k_yc_{y,\alpha}/|\omega| \ll 1$. Once satisfied for each beam, this latter criterion has been shown to grant an identical description of the linear WI-CFI modes in the kinetic and in the extended fluid models \citep{SarratEPL}.

Here we show that, in this limit, the kinetic TSI and the TRWI dispersion relations are correctly described by the extended fluid model.

For the sake of simplicity and in order to allow an analytical treatment of the transition between non time-resonant and time-resonant Weibel modes discussed in Section \ref{sec:4}, we restrict the analysis to the case of symmetric beams (Eqs.(\ref{eq:tsiconf_Dxx_DLposition}-\ref{eq:tsiconf_Dyy_DLposition})).
In this case, the hydrodynamic criterion reads
\begin{equation}
	\epsilon_\pm \equiv \frac{k_yc_y}{\sqrt{(\omega_r \mp k_yu_0)^2+\gamma^2}} \ll 1.
	\label{eq:criterion_sym}
\end{equation}

The equivalence between the kinetic and extended fluid linear analysis in the hydrodynamic limit, discussed in Sections \ref{sec:3.1} and \ref{sec:3.2} below, can be shown to hold also for non-symmetrical beams. In this latter case, it is usually more difficult to fulfil the hydrodynamic criterion simultaneously for the two beams, especially if the temperature asymmetry between the beams is strong. 

\subsection{Two-Stream Instability}\label{sec:3.1}
For the symmetrical TSI, the kinetic dispersion relation $D_{yy} = 0$ is \cite{LazarTAJ}
\begin{equation}
	0 = 1 + \frac{\omega^2_{pe}}{2}\left( \frac{1+\xi_+ Z_+}{k_y^2c^2_y} +\frac{1+\xi_- Z_-}{k_y^2c^2_y} \right),
	\label{eq:kin_TSI}
\end{equation}
where $Z_\pm = Z(\xi_\pm)$ is the plasma dispersion function \cite{Fried&Conte}, with $\xi_+ \equiv w_+/(\sqrt{2}k_y c_{y}) = (\omega - k_yu_0)/(\sqrt{2}k_y c_{y})$ and $\xi_- \equiv w_-/(\sqrt{2}k_y c_{y}) = (\omega + k_yu_0)/(\sqrt{2}k_y c_{y})$.

As the symmetrical TSI is a non-propagative instability ($\omega_r = 0$ case in Eq.(\ref{eq:criterion_sym})), $\epsilon_+ = \epsilon_-$ and so the hydrodynamic criterion is fulfilled by the two beams. Making the assumption that $\xi_\pm \gg 1$, it becomes possible to develop the plasma dispersion function
\begin{equation}
	Z(\xi) \sim -\frac{1}{\xi}\left(1+\frac{1}{2\xi^2} + \frac{3}{4\xi^4}\right),
	\label{eq:Zdev}
\end{equation}
and, substituting (\ref{eq:Zdev}) in (\ref{eq:kin_TSI}), we find
\begin{equation}
	0 \sim 1 - \frac{\omega^2_{pe}}{2} \left[\frac{1}{w_+^2}\left(1+\frac{3k_y^2c_y^2}{w_+^2}\right) + 
	\frac{1}{w_-^2}\left(1+\frac{3k_y^2c_y^2}{w_-^2}\right) \right].
	\label{eq:TSI_hydrodynamic}
\end{equation}
Such a development is justified because of the behaviour of the TSI : the parallel thermal spread decreases the growth rate of this instability. One has to retain the term of order four in $\xi$ in the development of $Z$ to preserve some thermal effects in the dispersion relation. The final contribution of this term is only of order $\xi^{-2}$.\\

If we perform a Taylor expansion of the the fluid dispersion relation (\ref{eq:tsiconf_Dyy_DLposition}) for $\epsilon_\alpha \ll 1$, we recover equation (\ref{eq:TSI_hydrodynamic}) implying the equivalence of the fluid and the kinetic descriptions when $\epsilon_\alpha \ll 1$ simultaneously for the two beams. 

\subsection{Weibel-like modes}\label{sec:3.2}

The kinetic dispersion relation $D_{xx}=0$ (or $D_{zz}=0$) for the symmetrical Weibel-like modes is \cite{LazarTAJ}
\begin{equation}
	\frac{k_y^2c^2}{\omega^2}=1+\frac{\omega_{pe}^2}{\omega^2}\left(A+\frac{A}{2}\left(\xi_+ Z_++\xi_- Z_-\right)-1\right),
	\label{eq:kin_weibel}
\end{equation}
where we define the anisotropy parameter $A \equiv c_x^2/c_y^2$.\\
Again we assume $\xi_\pm \gg 1$ and obtain the hydrodynamic dispersion relation
\begin{equation}
	\frac{k_y^2c^2}{\omega^2} \simeq 1-\frac{\omega_{pe}^2}{\omega^2}\left[1+\frac{A}{2}\left(\frac{k_y^2c_y^2}							{w_+^2}+\frac{k_y^2c_y^2}{w_-^2}\right)\right].
	\label{eq:weibeltsi_hydro}
\end{equation}
We have neglected all the termsgreater than $\epsilon_\pm^2$. This dispersion relation coincides with the corresponding limit taken for the fluid dispersion relation (\ref{eq:tsiconf_Dxx_DLposition}).

\section{Time-Resonant Weibel Instability}\label{sec:4}

From now on we focus on the dispersion relation of Weibel-like modes, $D_{xx} = 0$. The numerical analysis of the kinetic dispersion relation performed by \cite{LazarJPP} already evidenced the existence of a critical wave number $k_{_{SB}}$ at which a ``slope breaking'' of the growth rate occurs (\textit{cf.} Figure \ref{fig:2}), and the non-propagative nature of the most unstable mode existing for $k_y < k_{_{SB}}$. Here we show that the extended fluid approach allows an accurate description of such a transition and, moreover, makes it possible to identify the features of six roots of the dispersion relation and in particular of the second non-propagative mode met for $k_y < k_{_{SB}}$ : the slope-breaking is then found to correspond to a bifurcation point, at which the two couples of non-propagative unstable modes existing for $k_y < k_{_{SB}}$ merge into the four time-resonant growing and decaying modes found for $k_y > k_{_{SB}}$. The two growing modes - and also the two decaying ones - propagate in opposite directions (Figure \ref{fig:3}).

A second bifurcation point is found at the critical wave-number $k_y = k_c$ at which the growth rate of the unstable modes becomes zero (Figure \ref{fig:3}) : for $k_y > k_c$, all modes have now become purely propagative. Two former unstable modes propagate in each direction, with different values of $\omega_r$.

In order to show this, let us first discuss the accuracy of the fluid description in reproducing the known kinetic results (Section \ref{sec:4.1}). Then we will focus on the resolution of the fluid dispersion relation, which allows an analytic solution from which these results can be deduced (Section \ref{sec:4.2}). Then, in Section \ref{sec:4.3} we comment about the validity of these results in the kinetic regime.

\subsection{Transition to the time-resonant WI in the fluid and kinetic description}\label{sec:4.1}

\begin{figure}
\centerline{\includegraphics[width=14.5cm]{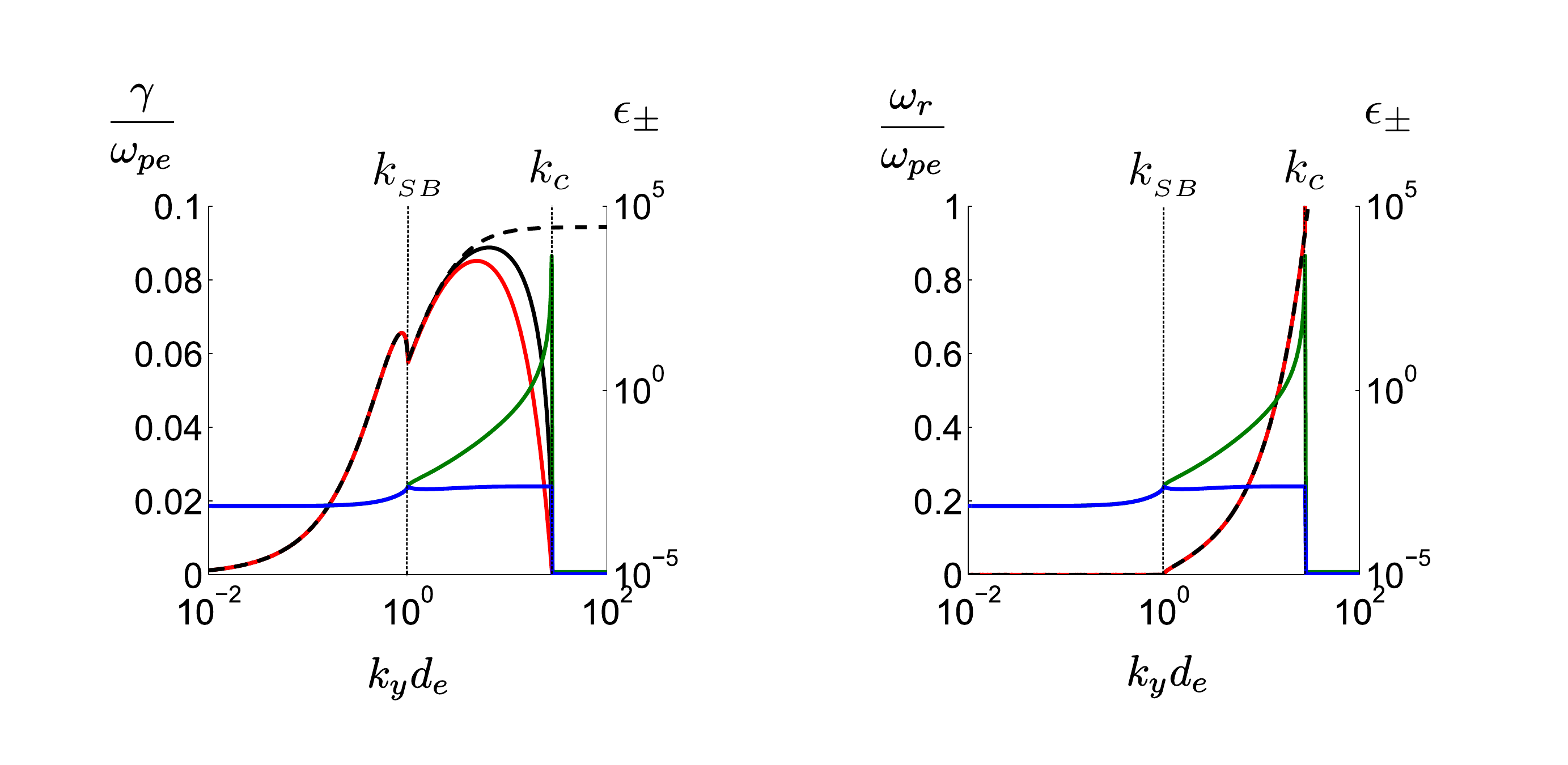}}
\caption{Comparison between kinetic (red line) and fluid (black line) growth rates (left panel) and corresponding frequencies (right panel) of the most unstable modes. The black dotted line corresponds to the values of $\omega_r$ and $\gamma$ obtained with equation (\ref{eq:weibeltsi_hydro}), \textit{i.e.} by applying the hydrodynamic limit. Differences between fluid and kinetic models are more important for the growth rate than for the frequency of the more unstable solution. On the right panel, the three curves overlap. Blue and green lines corresponds to the value of $\epsilon_\pm$, computed with the kinetic value of $\omega$ for the two beams (the green curve being for $\epsilon_-$, \textit{cf} Eq.\ref{eq:criterion_sym}) Physical parameters : $u_0 = 1/30$, $c_x = u_0 + 0.1$ and $c_y = u_0/10$.} 
\label{fig:2}
\end{figure}

Left panel of Figure \ref{fig:2} displays a comparison between the maximal growth rates obtained by integrating the fluid model, the kinetic model (both with no approximation) and the hydrodynamic limit of the two. The corresponding real frequencies, numerically obtained from equations (\ref{eq:tsiconf_Dxx_DLposition}) and (\ref{eq:kin_weibel}), are sketched in the right panel. We used for this a set of non-relativistic parameters : $u_0 = 1/30$, $c_x = u_0 + 0.1$ and $c_y = u_0/10$. Two points shall be highlighted.

About the discrepancies between the fluid and kinetic description,  one remarks a very good agreement between fluid and kinetic values of $\omega_r$ : differences between the two models and their hydrodynamic limit are negligible, even when $\epsilon_\pm$ is not. This is not the case of the growth rate, for which discrepancies between fluid and kinetic values increase with the wave number. However, they tend to coincide again at the cut-off value $k_c$, almost identical in the fluid and kinetic  model, next to which the hydrodynamic limit fails (the cut-off wave number disappears, instead, in the hydrodynamic limit (Eq.(\ref{eq:weibeltsi_hydro}))). The same kind of behaviour was remarked for the pure WI and for the WI-CFI coupled modes \cite{SarratEPL}.

Concerning instead the dependence of $\omega_r$ and $\gamma$ on the wave-number $k_y$, for the set of parameters of Figure \ref{fig:2} (right panel) one remarks the increase of $\omega_r$ between  $k_{_{SB}}d_e \sim 1$ and $k_cd_e \sim 30$. As evidenced by \cite{LazarJPP}, this behaviour is characteristic of the time-resonant WI, whose transition  at $k_{_{SB}}d_e \sim 1$ occurs in Figure \ref{fig:2} in the ``hydrodynamic" regime (both beams have $|\epsilon_\pm| \ll 1$ around $k_y = k_{_{SB}}$). The curves of the three growth rates exhibit the same slope breaking at $k_{_{SB}}$, indicating that for these parameters the pressure-tensor based model gives an accurate description of the transition.

Therefore, besides being accurate for values of $k_y$ smaller than $k_c$, yet comparable or larger than $k_{_{SB}}$, the linear analysis in the extended fluid model results reliable also outside the hydrodynamic limit, though with some quantitative differences with respect to the kinetic result. To study the behaviour of the modes given by the roots of Equation (\ref{eq:kin_weibel}), we then rely on the fluid dispersion relation (\ref{eq:tsiconf_Dxx_DLposition}), which has the advantage of allowing a simpler analytical treatment due to its polynomial form in $\omega$ and in $k_y$.

\subsection{Discussion of the fluid dispersion relation}\label{sec:4.2}
The fluid dispersion relation (\ref{eq:tsiconf_Dxx_DLposition}) gives a polynomial of degree three in $\omega^2$,
\begin{eqnarray}
0 & = & \omega^6-\omega^4\big[\omega_{pe}^2+k_y^2c^2+2k_y^2(u_0^2+c_y^2)\big]\nonumber\\
&& +\;\omega^2\big[k_y^4(u_0^2-c_y^2)^2 + 2k_y^2(u_0^2+c_y^2)(\omega_{pe}^2+k_y^2c^2)-\omega_{pe}^2k_y^2c_x^2\big]\nonumber\\
&& -\;k_y^4\big(c_y^2-u_0^2\big)\big[k_y^2c^2(c_y^2-u_0^2)-\omega_{pe}^2(u_0^2+c_x^2-c_y^2)\big].
\label{eq:weibeltsi_polynomial}
\end{eqnarray}

\begin{figure}
\centerline{\includegraphics[width=14.cm]{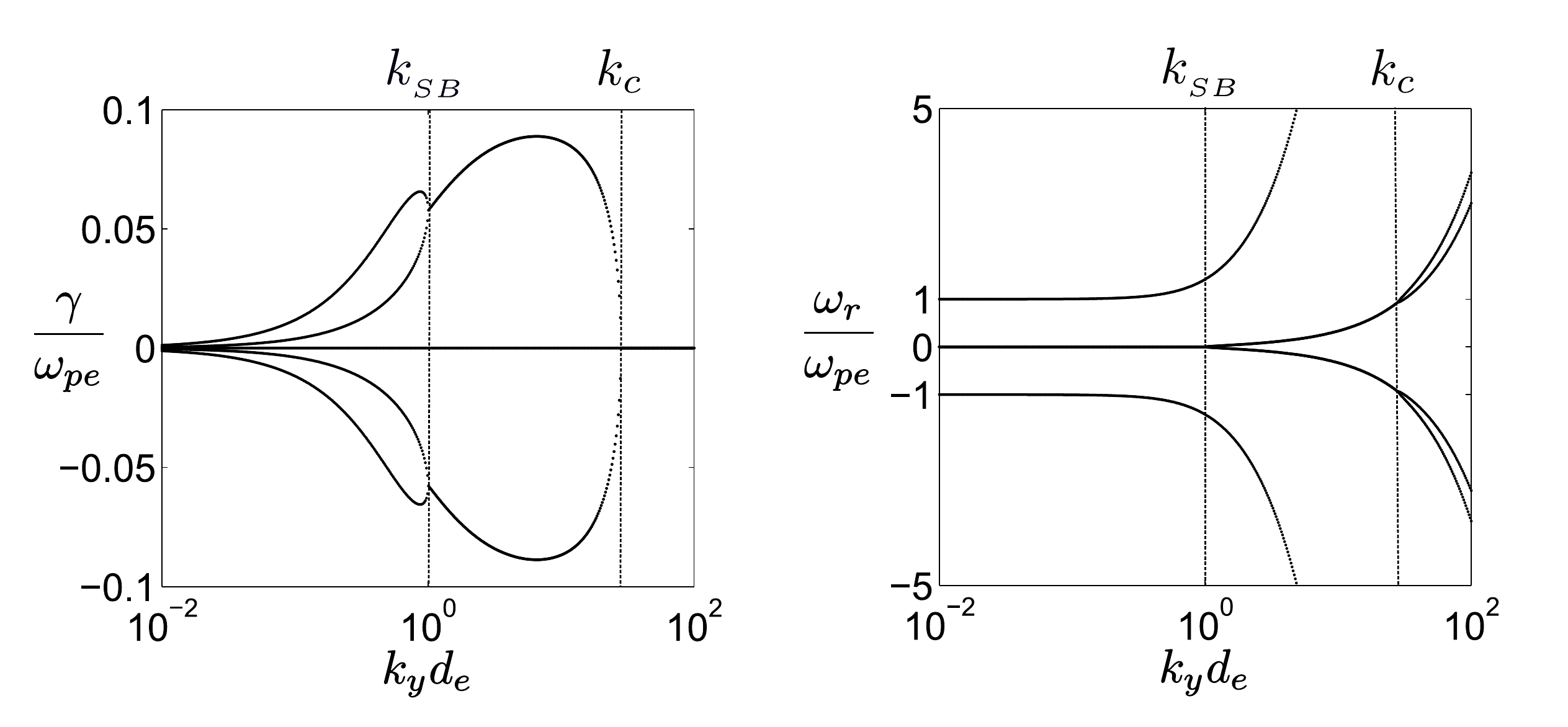}}
\caption{Roots of the equation (\ref{eq:weibeltsi_polynomial}) : growth rates (left panel), frequencies (right panel). The two panels display, for any value of $k_y$, the behaviour of a total of six modes.  The slope-breaking observed in Figure \ref{fig:2} for the growth rate corresponds to a merging of two unstable modes. Two other modes (characterised by $\omega_r \geq \omega_{pe}$) are stable for any value of $k_y$. Physical parameters : $u_0 = 1/30$, $c_x = u_0 + 0.1$ and $c_y = u_0/10$.} 
\label{fig:3}
\end{figure}

which implies the existence of three couples of complex roots. These roots are evidenced in Figure \ref{fig:3}, where a numerical solution of Equation (\ref{eq:weibeltsi_polynomial}) is provided for the same parameters of Figure \ref{fig:2}.

Although it is in principle possible to write the explicit general form of the roots of (\ref{eq:weibeltsi_polynomial}), for the purpose to identify the propagative or unstable behaviour of the modes in the intervals $k_y \leq k_{_{SB}}$, $k_{_{SB}} \leq k_y \leq k_c$ and $k_y > k_c$, it is sufficient to discuss the existence of the real and imaginary parts of the $\omega^2$ solutions. Rewriting (\ref{eq:weibeltsi_polynomial}) in a more compact form,
\begin{equation}
	W^3 + a_2 W^2 + a_1 W + a_0 = 0,
\end{equation}
with $W \equiv \omega^2$, we look at the sign of \cite{Abramovitz}
\begin{equation}
	\Delta \equiv q^3 + r^2,
	\label{eq:discr6}
\end{equation}
where
\begin{equation}
	q = \frac{1}{3}a_1 - \frac{1}{9}a_2^2 , \quad r = \frac{1}{6}(a_1a_2-3a_0)-\frac{1}{27}a_2^3.
	\label{eq:qr}
\end{equation}
When $\Delta > 0$, one root is real and two are complex conjugate. In the opposite case $\Delta < 0$, the irreducible case, there are three different real solutions, while if $\Delta = 0$, all the roots are real and at least two of them coincide.

The most general expressions for the three solutions in $\omega^2$ read
\begin{equation}
	\omega^2_{0} = \underbrace{[r+\Delta^{\frac{1}{2}}]^{\frac{1}{3}}}_{\displaystyle{s_+}} + \underbrace{[r-\Delta^{\frac{1}{2}}]^{\frac{1}{3}}}_{\displaystyle{s_-}} - \frac{a_2}{3}
	\label{eq:deg6_1}
\end{equation}
and
\begin{equation}
	\omega^2_{\pm} = -\frac{1}{2}(s_+ + s_-) - \frac{a_2}{3} \pm \frac{i\sqrt{3}}{2}(s_+-s_-)
	\label{eq:deg6_2&3}
\end{equation}

As $W = \omega^2 = \omega_r^2 - \gamma^2 + i2\omega_r \gamma$, the only possibility to obtain a real $\omega^2$ is $\omega_r = 0$ or $\gamma = 0$, that is, to have purely propagative or purely unstable modes, whereas an unstable propagative mode corresponds to a complex value of $\omega^2$. This allows to characterise in terms of the sign of $\Delta$ the three regions in the $k_y$ parameter-space identified above and evidenced in the example of Figure \ref{fig:3}. Moreover, as the discriminant $\Delta$ is a continuous function of $k_y$, the slope-breaking $k_{_{SB}}$ and the cut-off wave number $k_c$ correspond to the roots of the equation $\Delta (k_y) = 0$. This gives a criterion to determine \textit{a priori} whether the WI is time-resonant or strictly growing - or damped. Let us discuss under this light the three regions of Fig \ref{fig:3}.

$k_y < k_{_{SB}}$ region : here the discriminant $\Delta$ is negative, assuring three strictly different real values for $\omega^2$, either purely propagative of purely unstable. The two stable modes propagating in opposite directions and characterised by $|\omega_r| \geq \omega_{pe}$ correspond to linearly polarised electromagnetic waves. The four other modes consist of two non-propagative and growing Weibel-type modes and of their two damped counterparts, which evolve on long time scales with respect to the inverse electron oscillation time, $\gamma \ll \omega_{pe}$. For the set of parameters of Figure \ref{fig:3}, a local maximum in the growth rates can be identified around $k_yd_e \sim 0.8$.

$k_y = k_{_{SB}} \sim d_e^{-1}$ represents a ``double'' bifurcation point in the $(k_y,\omega)$ space, which solves $\Delta (k_y) = 0$. At the slope breaking of $\gamma$, first observed by \cite{LazarJPP}, the two purely growing (damped) roots, distinct for $k_y < k_{_{SB}}$, acquire the same growth rate and a propagative character, pairwise, in opposite directions.

$k_{_{SB}} < k_y < k_c$ region : here, where time-resonant unstable modes are encountered, $\Delta > 0$. The time-resonant Weibel modes, propagating in opposite ways (same $\gamma$ and opposite $\omega_r$), are described by the roots $\omega_+^2$ and $\omega_-^2$ (\ref{eq:deg6_2&3}). Moreover, damped and growing modes are complex conjugate. The stable electromagnetic waves already encountered for $k_y < k_{_{SB}}$ are given by the only real root $W = \omega_0^2$ (\ref{eq:deg6_1}).

$k_y = k_c$ is the point at which the unstable part of the solutions vanish. It corresponds again to a ``double'' bifurcation in the $(k_y,\omega)$ space, once more given by $\Delta(k_c) = 0$.

In the $k_y > k_c$ region, where $\Delta < 0$ again, all modes maintain their propagative character with two different couples of values of $\omega_r$, pairwise opposite in sign. All modes propagating above this critical wave-number are stable, linearly polarised electromagnetic waves.\\

\begin{figure}
\centerline{\includegraphics[width=12.cm]{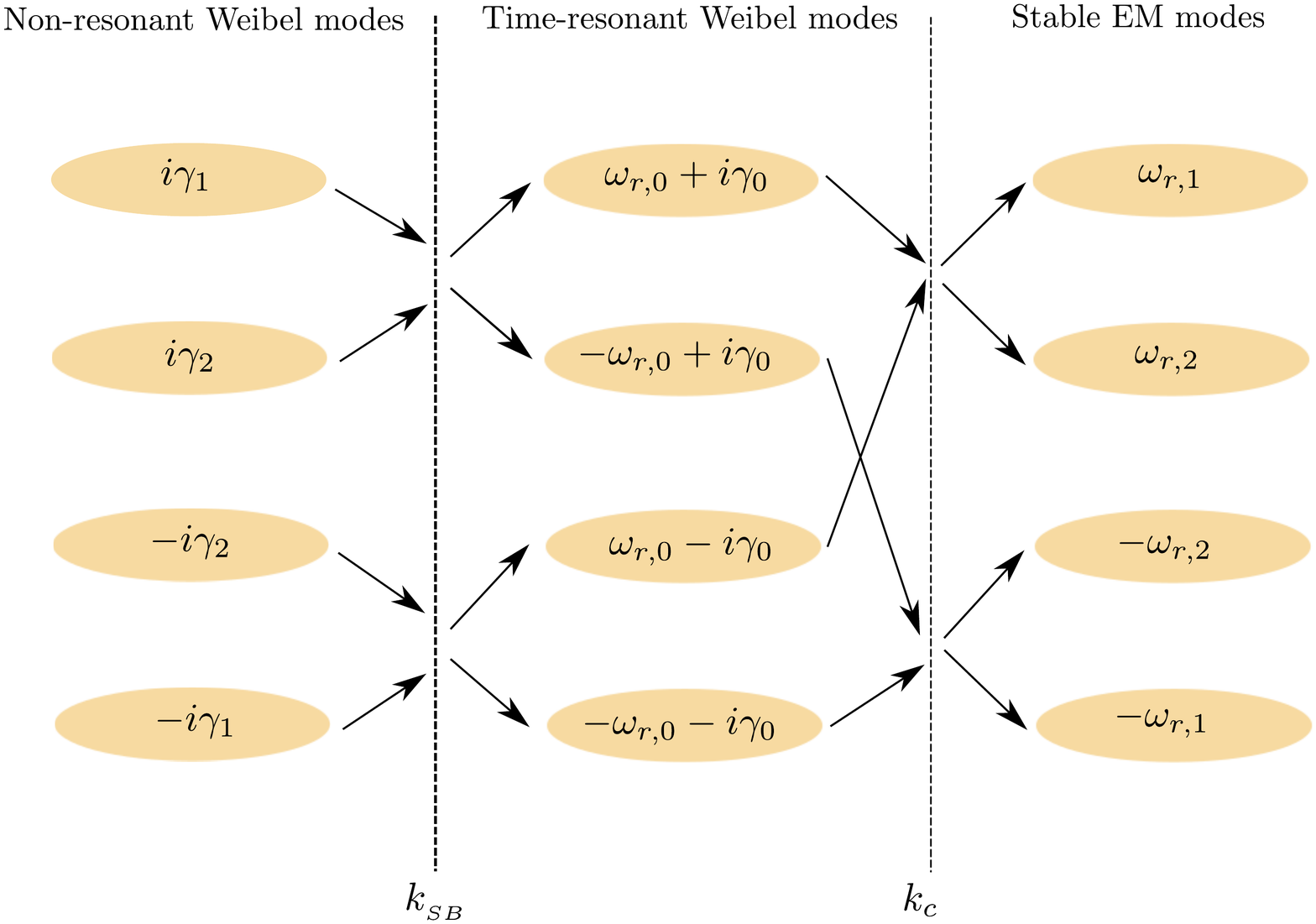}}
\caption{Complex general expression of the four unstable modes as a function of the wave number. Black arrows indicate a merging or a breaking of degeneracy of two modes at a bifurcation. The bifurcations, at $k_y = k_{_{SB}}$ and $k_y = k_c$, are indicated by the dashed vertical lines. Here, subscripts $\alpha = 1,2$ generically label differrent values for $\gamma$ and $\omega_r$.} 
\label{fig:recap}
\end{figure}

The nature of the six fluid modes in these three regions of the wave-number space is summarised in Figure \ref{fig:recap}.

\subsubsection{Low frequency approximation}\label{sec:4.2.1}

Further insight on the behaviour of the unstable modes comes from the restriction to low frequencies $|\omega| \ll \omega_{pe}$, which implies the exclusion of the stable solutions and thus the reduction of Equation (\ref{eq:weibeltsi_polynomial}) to a polynomial of degree two in $\omega^2$ :
\begin{eqnarray}
0 & = & \omega^4\left[\omega_{pe}^2+k_y^2c^2\right] - \omega^2\left[2k_y^2(u_0^2+c_y^2)(\omega_{pe}^2+k_y^2c^2)-\omega_{pe}^2k_y^2c_x^2\right]\nonumber\\
&& +\;k_y^4\left(c_y^2-u_0^2\right)\left[k_y^2c^2(c_y^2-u_0^2)-\omega_{pe}^2(u_0^2+c_x^2-c_y^2)\right].
\label{eq:deg6_ordering1}
\end{eqnarray}

It turns out that the growth rate and the real frequency of the four modes remain practically unchanged under this approximation. For example, for the parameters used in Figure \ref{fig:2}, the maximal growth rate in the non-resonant and in the time-resonant regime undergo a variation of $\sim 0.1\%$, whereas $k_c$ and $k_{_{SB}}$ vary less than $0.1\%$.\\

With similar arguments to those previously developed to analyse the roots of Equation (\ref{eq:weibeltsi_polynomial}), we obtain quite accurate analytical estimations of the slope-breaking and cut-off wave numbers. We thus look for the roots of $\Delta_*(k_y) = 0$,  where $\Delta_*$ is the discriminant of Eq. (\ref{eq:deg6_ordering1}), which reads
\begin{equation}
\Delta_* = k_y^4 \left\{ \left[16u_0^2c_y^2c^4\right]k_y^4 + 8\omega_{pe}^2u_0^2c^2\left[4c_y^2-c_x^2\right]k_y^2
+\;\omega_{pe}^4\left[c_x^4 + 8 u_0^2(2c_y^2-c_x^2)\right]\right\}.
\label{eq:deg4_discr}
\end{equation}

The roots of the above polynomial correspond to real values of $k_y^2$ when $c_y^2 < u_0^2$ (condition met for the set of parameters of Figure \ref{fig:3}). Under this hypothesis, the slope-breaking and the cut-off wave numbers respectively read :
\begin{equation}
	k_{_{SB}}^2 d_e^2 = \frac{1}{4}\left[ \frac{c_x^2}{c_y^2}\left(1 - \sqrt{1 - \frac{c_y^2}{u_0^2}} \right)-4\right], \qquad
	k_{c}^2 d_e^2 = \frac{1}{4}\left[ \frac{c_x^2}{c_y^2}\left(1 + \sqrt{1 - \frac{c_y^2}{u_0^2}} \right)-4\right].
	\label{eq:kSBkc_expr}
\end{equation}
For the parameters of Figure \ref{fig:3}, for example, the values $k_{_{SB}}=1.0025$ and $k_c = 28.2311$ are obtained, in good agreement with the numerical solution of Equation (\ref{eq:weibeltsi_polynomial}).

The analytical estimation of $k_{_{SB}}$ given by \cite{LazarJPP} in the hydrodynamic regime is recovered by taking the further limit $c_y \ll u_0$, for which (\ref{eq:kSBkc_expr}) becomes :
\begin{equation}
	k_{_{SB}} d_e \simeq \sqrt{\frac{c_x^2}{8u_0^2}-1} \quad; \quad k_{_c} d_e \simeq \frac{c_x}{\sqrt{2} c_y}.
	\label{eq:kSBkc_simp}
\end{equation}
The limit value $u_0 = c_y$ results particularly interesting as it implies the absence of the time-resonant solutions, due to the superposition of the roots
\begin{equation}
	k_{_{SB}}d_e = k_{_c}d_e = \sqrt{\frac{c_x^2}{4c_y^2}-1}.
	\label{Weibelcase_kc}
\end{equation}

Only the non-propagative Weibel modes remain. One remarks the strong similarity between (\ref{Weibelcase_kc}) and the cut-off wave-number of the pure WI modes, $k_c^{WI} = \sqrt{c_x^2/c_y^2 - 1}$, as for $c_y = u_0$ the sixth order polynomial dispersion relation (\ref{eq:weibeltsi_polynomial}) becomes
\begin{equation}
0 = \omega^4 - \omega^2 \left[\omega_{pe}^2+k_y^2(c^2+4c_y^2)\right]+k_y^2\left[k_y^2c^24c_y^2-\omega_{pe}^2(c_x^2-4c_y^2)\right].
\label{eq:weibeltsi_degree4}
\end{equation}

This represents, indeed, the pure WI dispersion relation as obtained in the fluid framework (cf. Eq.(22) of \cite{SarratEPL}), with an effective squared ``thermal'' velocity defined as $\tilde{c}_y^2 = 4c_y^2$. This specific point can be understood considering each beam as represented by a bi-Maxwellian distribution function, whose standard deviations in the velocity space are $\sigma_i = c_i$, with $i = x,y$ : when $c_y = u_0$, the two Maxwellian are so close that they shape one single Maxwellian with standard deviation $\widetilde{\sigma}_i = 2 \sigma_i$.

\begin{figure}
\centerline{\includegraphics[width=12.cm]{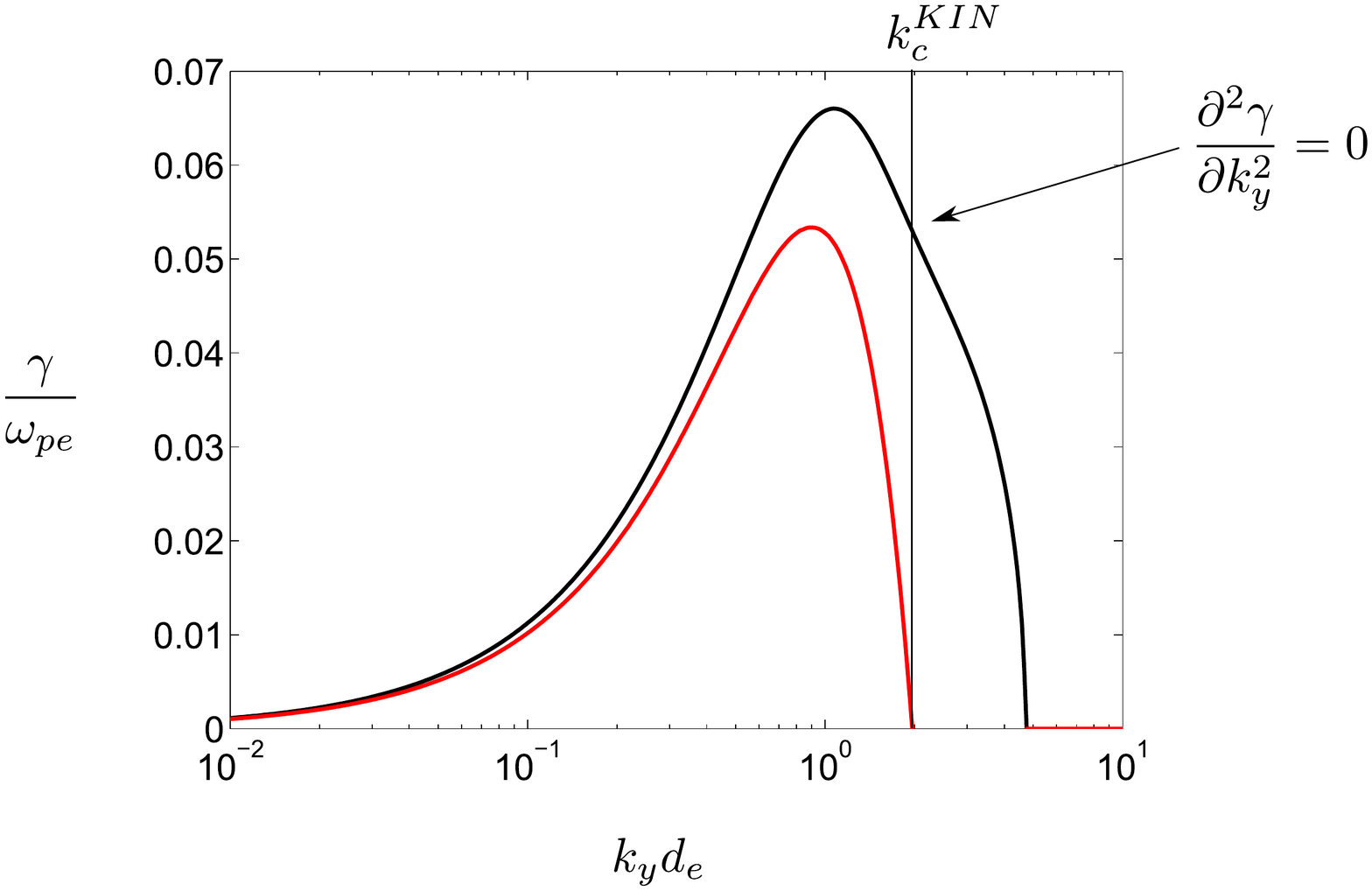}}
\caption{Comparison between fluid (black) and kinetic (red) growth rates for $u_0 = 1/30$, $c_y = u_0 + 0.01$ and $c_x = u_0 + 0.1$. The spurious inflection point on the fluid growth rate is observed when $k_y$ is close to the kinetic cut-off $k_c^{KIN}$.} 
\label{fig:cygtu0}
\end{figure}

If $c_y > u_0$ the  two distribution functions overlap and a part of the information contained in each of them is lost. We can then expect that these non-resonant Weibel modes will persist in the fluid description  until $\widetilde{c}_y$ becomes comparable to $c_x$. A comparison between fluid and kinetic solutions (Figure \ref{fig:cygtu0}) shows that both  descriptions lead to a non-resonant instability. However, the fluid growth rate exhibits an inflection point around the kinetic cut-off which does not exist in the kinetic description : once more, this suggests that close to the kinetic cut-off wave number, the inclusion of higher order velocity moments in the fluid model becomes necessary.

\subsubsection{Low frequency, small wave numbers limit}\label{sec:4.2.2}

A simplified analytical expression for the growth rates of the non-resonant Weibel modes  existing for $k_y < k_{_{SB}}$ can be obtained by further considering the $k_yd_e \ll 1$ limit of Equation (\ref{eq:deg6_ordering1}) :
\begin{equation}
	\omega^4 - \omega^2\left[2k_y^2(u_0^2+c_y^2)-k_y^2c_x^2\right] + k_y^4(c_y^2-u_0^2)(c_y^2-u_0^2-c_x^2) = 0.
\end{equation}
Its discriminant reads :
\begin{equation}
	\Delta_{**} = k_y^4c_x^4\left[1 + \frac{8u_0^2}{c_x^2}\left(\frac{2c_y^2}{c_x^2}-1\right)\right]
				= k_y^4 \left[c_x^4 + 8u_0^2(2c_y^2-c_x^2) \right]
\end{equation}
and its sign does not depends on the value of $k_y$. Consequently the bifurcations disappear.\\

\begin{figure}
\centerline{\includegraphics[width=8.cm]{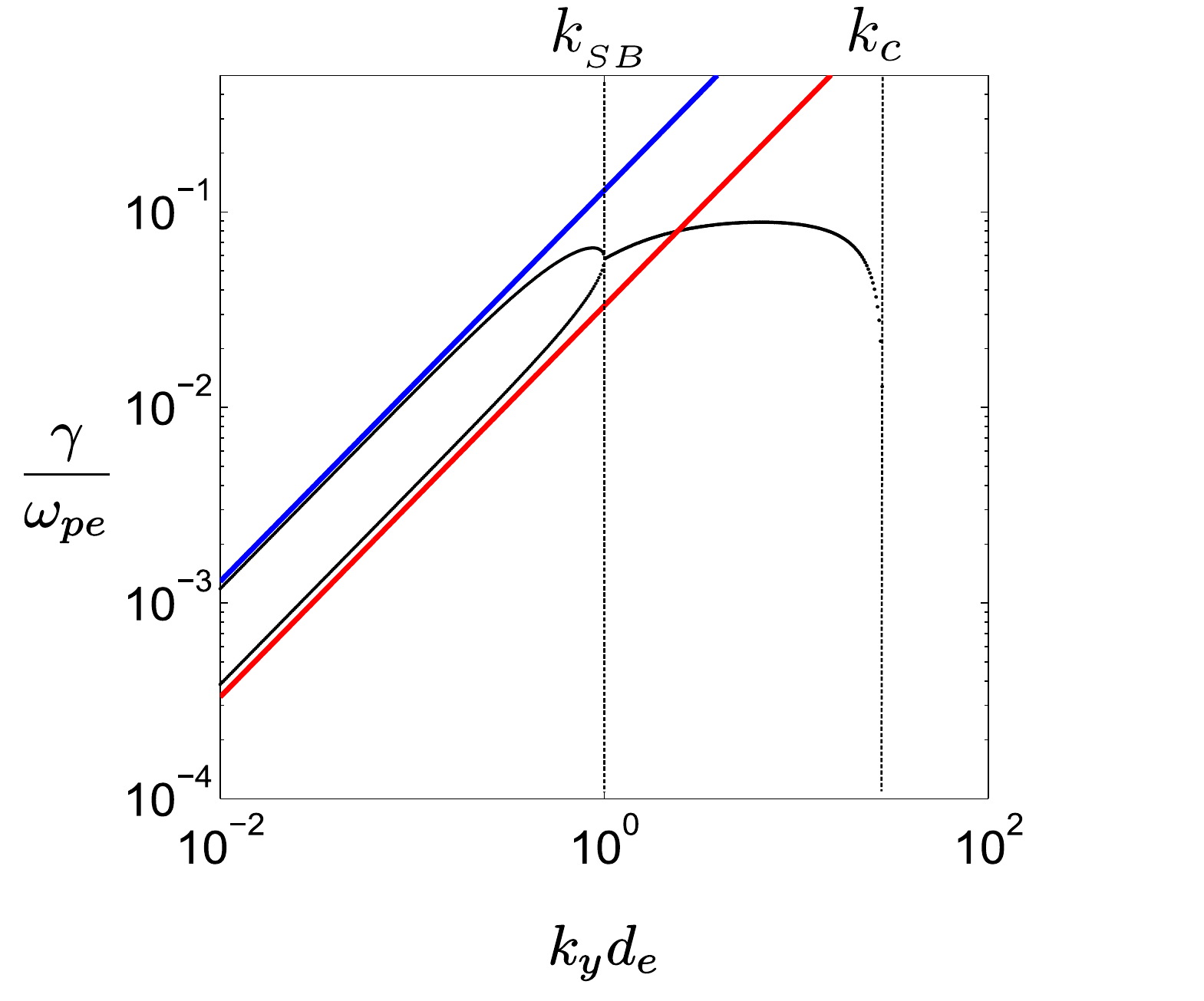}}
\caption{The two asymptotic solutions (blue and red lines) of the purely growing modes for $k_y d_e \ll 1$. Same parameters of Figures \ref{fig:2} and \ref{fig:3}.} 
\label{fig:asymp}
\end{figure}

Consider now the strong anisotropy limit expressed by $c_y \ll u_0 \ll c_x$, which is, for example, the one represented in Figures \ref{fig:2} and \ref{fig:3} (with $\Delta_{**}$ scaling as $10^{-4}k_y^4$). This corresponds to the case $\Delta_{**} > 0$. This assumption leads us to the following asymptotic estimates for the two solutions in $\omega^2$ and for the corresponding growth and damping rates (Figure \ref{fig:asymp}),
\begin{equation}
	\omega^2 \simeq - k_y^2(u_0^2-c_y^2), \qquad\gamma \simeq \pm k_y u_0 \sqrt{1 - \frac{c_y^2}{u_0^2}}\qquad \text{(least unstable mode),}
	\label{eq:asyp_nrs} 
\end{equation}
\begin{equation}
	\omega^2 \simeq - k_y^2(c_x^2-3u_0^2), \qquad \gamma \simeq \pm k_y c_x \sqrt{1 - \frac{3u_0^2}{c_x^2}}\qquad \text{(most unstable mode),}
	\label{eq:asym_nrs} 
\end{equation}
which, as expected, are non propagative.\\

Two time-resonant unstable modes can be obtained instead for $\Delta_{**} < 0$. Since the discriminant of $\Delta_{**}(c_x^2) = 0$ is $64u_0^2(u_0^2-c_y^2)$, we deduce that in this case $\Delta_{**} < 0$ if $c_{x-}^2 < c_x^2 < c_{x+}^2$, with
\begin{equation}
	c_{x,\pm}^2 = 4u_0^2\left( 1 \pm \sqrt{1-\frac{c_y^2}{u_0^2}} \right).
	\label{eq:alwres_bound}
\end{equation}

Assuming in particular, with no loss of generality, $c_y \ll u_0$ then $c_{x,+} \simeq \sqrt{2(4u_0^2-c_y^2)} \simeq 2\sqrt{2}u_0$ and $c_{x,-} \simeq \sqrt{2}c_y$. This result agrees with (\ref{eq:kSBkc_simp}) for the slope-breaking : for $c_y \ll u_0$, when $c_x < 2\sqrt{2}u_0$ the Weibel instability is always propagative.\\

\begin{figure}
\centerline{\includegraphics[width=14.cm]{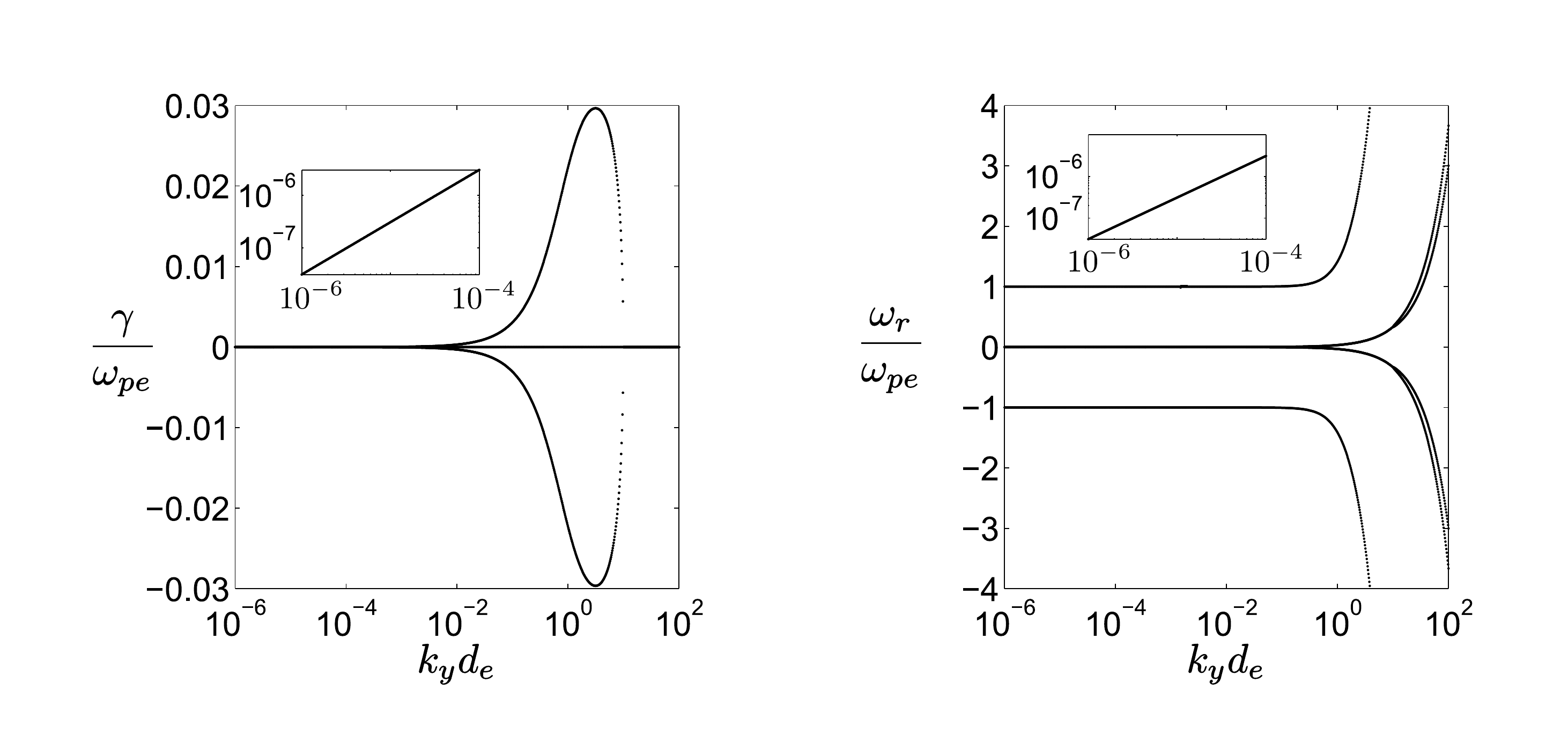}}
\caption{Numerical solution of the fluid dispersion relation (Eq. \ref{eq:weibeltsi_polynomial} for $u_0 = 1/30$, $c_y = u_0/10$ and $c_x = \sqrt{2}u_0$ (then $\Delta_{**} \simeq -1,5.10^{-5}k_y^4$).} 
\label{fig:421}
\end{figure}

To illustrate this point, let us consider a numerical example for parameters fitting in the upper bound of the interval $\Delta_{**} < 0$ : choosing $u_0 = 1/30$, $c_y = u_0/10$, which fulfil the condition $c_y \ll u_0$, and $c_x = \sqrt{2} u_0 \lesssim c_{x,+}$, then $\Delta_{**} \simeq -1,5.10^{-5}k_y^4$. The solutions of the exact fluid dispersion relation (\ref{eq:weibeltsi_polynomial}) are displayed in Figure \ref{fig:421}. Note that, because of this, the assumptions $k_yd_e \ll 1$ and $|\omega|\ll \omega_{pe}$ do not intervene in this result. Looking at the sign of   the discriminant $\Delta$ of the exact fluid dispersion relation (\ref{eq:weibeltsi_polynomial}), the represented curve is recognised to correspond to the TRWI, with a value of $k_c$ which is well approximated by the corresponding estimation of (\ref{eq:kSBkc_simp}) : $\Delta$ is positive for $k_y < k_c \sim 10 d_e^{-1}$, zero at $k_y = k_c$ and negative for $k_y > k_c$.\\

\begin{figure}
\centerline{\includegraphics[width=14.cm]{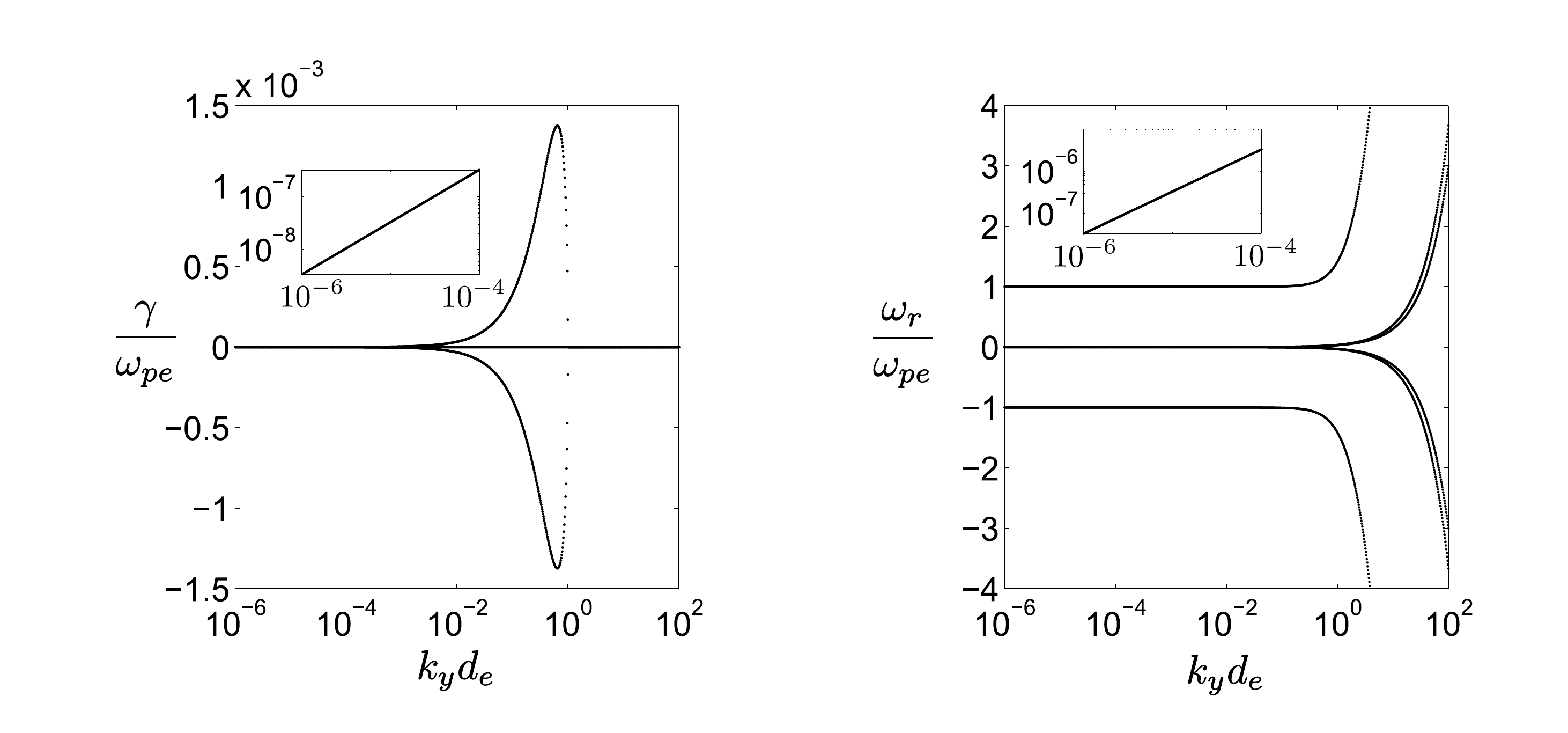}}
\caption{Numerical solution of the fluid dispersion relation (Eq. \ref{eq:weibeltsi_polynomial} for $u_0 = 1/30$, $c_y = u_0/10$ and $c_x = 2c_y$ (then $\Delta_{**} \simeq -2.10^{-7}k_y^4$).} 
\label{fig:422}
\end{figure}

Let us focus now on the lower bound of the $\Delta_{**} < 0$ interval, by choosing $u_0 = 1/30$, $c_y = u_0/10$ and $c_x = 2c_y$, so that $\Delta_{**} \simeq -2.10^{-7}k_y^4$. Results are shown in Figure \ref{fig:422}. As expected, due to the smaller thermal anisotropy, the growth rate results smaller than the one considered in Figure \ref{fig:421}, but the qualitative behaviour is the same.\\

Finally, $\Delta_{**} > 0$  and  there are no more unstable modes when  $c_x \lesssim c_{x,-}$ because the transverse thermal energy is not sufficient anymore to drive the Weibel-type mode.

\subsection{Full kinetic description}\label{sec:4.3}

Although with some quantitative differences in the numerical estimations of the growth rates and critical wave numbers, the existence and characterisation of the sixth roots of the fluid dispersion relation (\ref{eq:weibeltsi_polynomial}) in the three intervals of the wave number space, $k_y < k_{_{SB}}$, $k_{_{SB}} < k_y < k_c$ and $k_c < k_y$, is confirmed in a full kinetic description, even when the hydrodynamic limit is not satisfied. In particular, the second unstable mode evidenced in Figure \ref{fig:3}  can be generally identified also in the kinetic framework (we note en passing the usefulness of the fluid solution as an initial seed for the numerical kinetic solver).

\begin{figure}
\centerline{\includegraphics[width=14.cm]{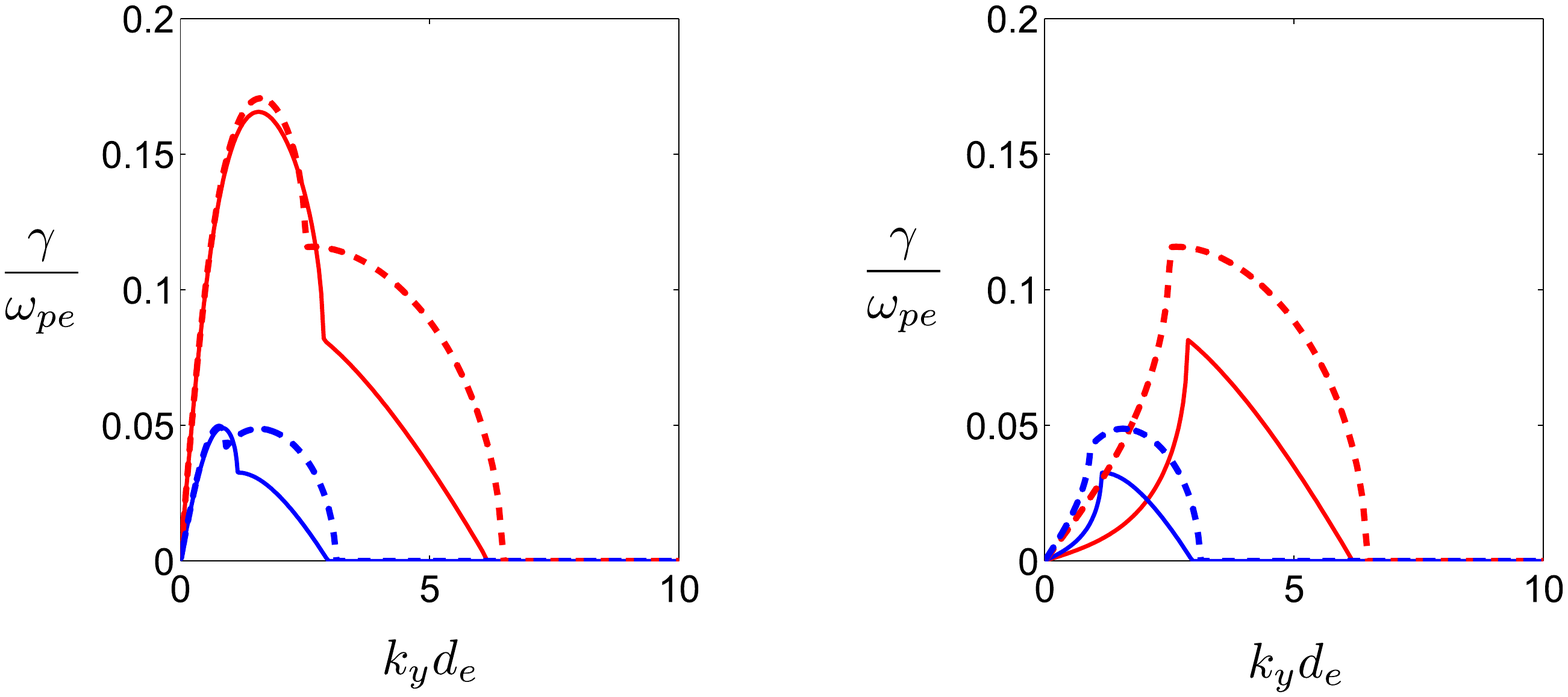}}
\caption{Comparison between fluid (dashed lines) and kinetic (solid lines) growth rates corresponding to the upper (left panel) and to the lower (right panel) unstable branches of the Weibel instability. Blue curves are obtained for $u_0 = 1/30$, $c_y = u_0/\sqrt{2}$ and $c_x = 10c_y$. Red curves are obtained for $u_0 = 1/30$, $c_y = u_0/\sqrt{2}$ and $c_x = 5c_y$. A direct comparison with Figure 1 in \cite{LazarJPP} can be done.} 
\label{fig:432}
\end{figure}

An example of this solution is shown in Figure \ref{fig:432}, where the growth rate of the lower non-resonant Weibel unstable branch is represented as a function of the wave number, for parameters which do not satisfy the hydrodynamic criterion. The fluid model overestimates the kinetic growth rate, but $k_{_{SB}}$ is roughly the same. Note that due to the lower value of $|\omega|$, the hydrodynamic criterion is harder to achieve for the lower branch than for the upper branch.\\

\begin{figure}
\centerline{\includegraphics[width=14.cm]{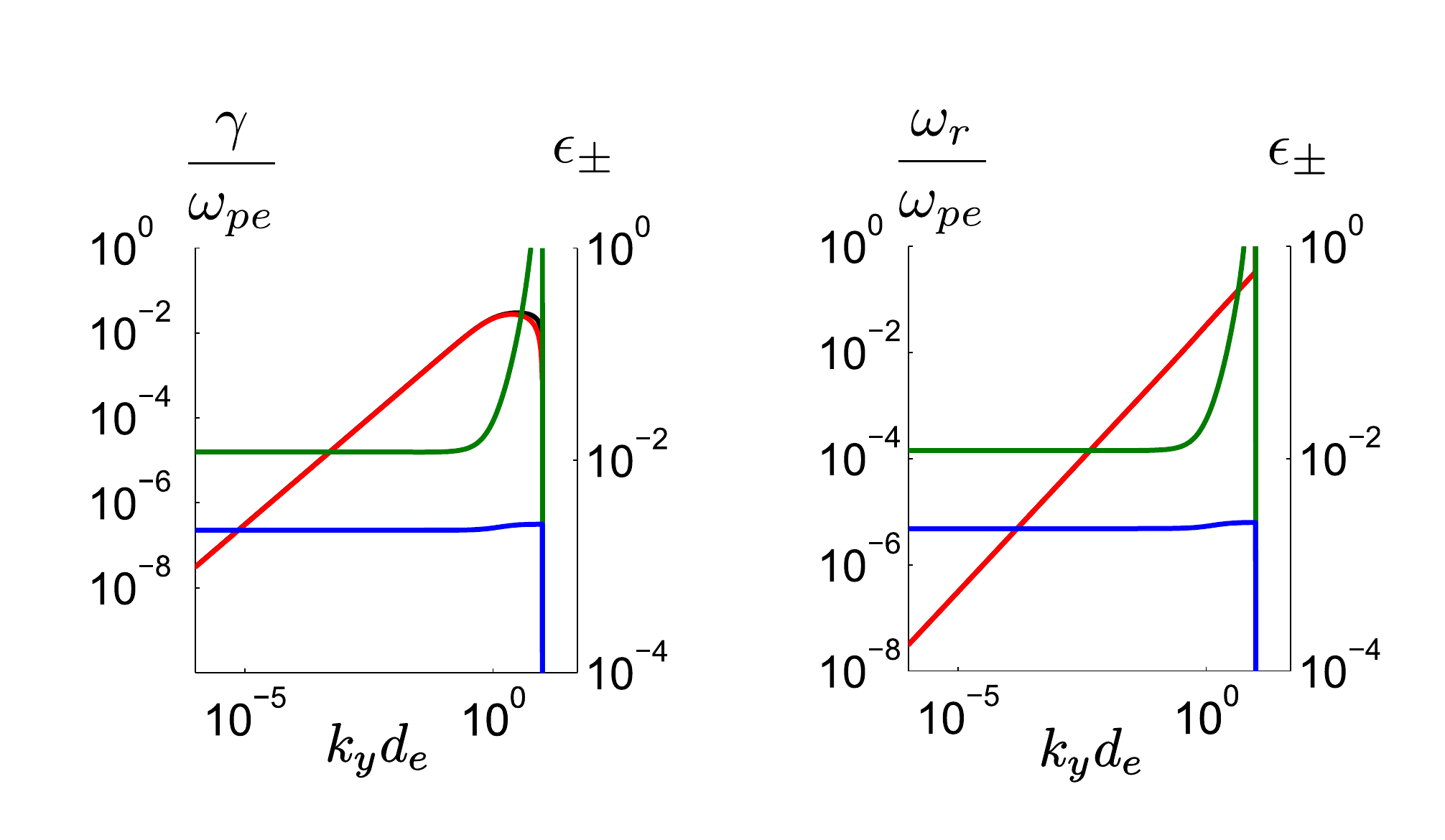}}
\caption{Comparison between kinetic (red line) and fluid (black line) maximal growth rates (left panel) and corresponding frequencies (right panel). Blue and green lines corresponds to the hydrodynamic criterion $\epsilon_\alpha = k_yc_y/\Omega_\alpha$ values for the two beams, computed with the kinetic value of $\omega$. The physical parameters are the same than for Figure \ref{fig:421} : $u_0 = 1/30$, $c_x = \sqrt{2}u_0$ and $c_y = u_0/10$. Differences between the two models appears only near of the cut-off.} 
\label{fig:433}
\end{figure}

\begin{figure}
\centerline{\includegraphics[width=16.cm]{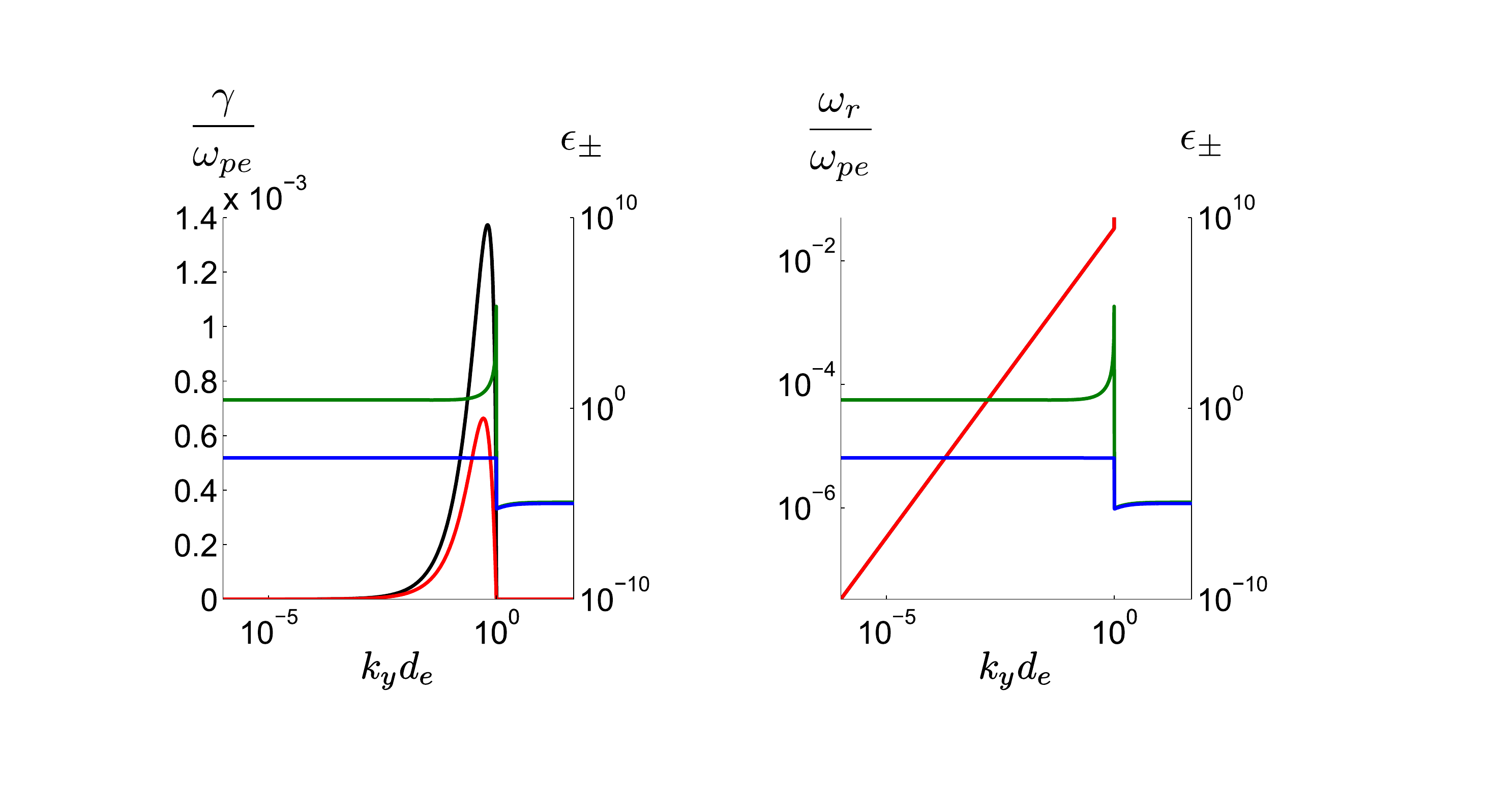}}
\caption{Comparison between kinetic (red line) and fluid (black line) maximal growth rates (on the left panel) and corresponding frequencies (on the right one). Blue and green lines corresponds to the value of the hydrodynamic criterion $\epsilon_\alpha$ for the two beams. The physical parameters are the same than for Figure \ref{fig:422} : $u_0 = 1/30$, $c_y = u_0/10$ and $c_x = 2c_y$. The hydrodynamic limit is never reached, but the two models predict a Resonant Weibel regime whatever $k_y$. If the differences between the growth rate values are substantial, the agreement between the two descriptions for the real frequency is very good.} 
\label{fig:434}
\end{figure}

The role of thermal effects on the transition to the TRWI, characterised in terms of the fluid analysis in Section \ref{sec:4.2.2}, is also recovered in the kinetic framework. As an example, in Figures \ref{fig:433} and \ref{fig:434} we show a comparison of the fluid and kinetic growth rates and real frequencies for the most unstable mode analysed in Figures \ref{fig:421} and \ref{fig:422} respectively.

For the case of figure \ref{fig:433} the agreement is excellent, coherently with the fulfillment of the hydrodynamic criterion for this set of parameters. We can expect, also in the kinetic case, the disappearance of the non-resonant instability in favour of the time-resonant one, when the condition $c_{x-}^2 < c_x^2 < c_{x+}^2$ is fulfilled, as the behaviour observed for  $k_yd_e \sim 10^{-6}$ is the same to that observed at arbitrarily smaller wave number. Indeed, the terms  $\sim k_y^8$ and $ \sim k_y^6$ in equation (\ref{eq:deg4_discr}) are dominated by those proportional to $k_y^4$.

Nevertheless, the qualitative agreement is excellent also for the case of Figure \ref{fig:434}, which is largely outside the hydrodynamic regime, though quantitative differences are important  here : the fluid model largely overestimates the kinetic value of the maximal growth rate.  It remains however correctly smaller than the fluid one displayed on Figure \ref{fig:433}, as expected due to the relatively smaller initial anisotropy. Remarkably, the fluid and kinetic estimations of $k_c$  and  of $\omega_r$ for  the TRWI appear to be in excellent quantitative agreement also in this case.

\section*{Conclusion}

In this work we have shown that an extended fluid description which takes into account the full pressure tensor dynamics makes it possible to reproduce important features of Weibel-type instabilities excited by perturbations parallel to two counterstreaming beams, such as the existence of time-resonant (propagating) modes. By restricting to the case of symmetric beams, we have evidenced an excellent quantitative agreement between the fluid and kinetic linear description of both the TSI and the Weibel-like modes in  the hydrodynamic limit (Section \ref{sec:3}).  We have then shown how the behaviour of the fluid modes agrees qualitatively and, partially, quantitatively, with the kinetic description also outside of this hydrodynamic regime  (Section \ref{sec:4}).\\

We have shown that this fluid description makes possible an analytical identification of various features  of these Weibel-like modes in the parameter space. For example, this approach has allowed to evidence a second purely unstable mode in the non-resonant regime, whose existence is attested also in the kinetic framework even when the hydrodynamic assumption is not satisfied. The slope-breaking in the growth rate, first observed by \cite{LazarJPP}, is then found to correspond to a bifurcation point of two unstable non-resonant modes, which completely disappears for a sufficiently small initial anisotropy (condition which leads, instead, to a purely TRWI $-$ Section \ref{sec:4.2.2}). \\

Future numerical investigations will tell whether the fluid description of these Weibel-type modes remains consistent also during their non-linear evolution, possibility which would imply a remarkable gain in terms of computational cost of nonlinear simulations. In this regard we remark the interest of this analysis also as a further step towards the fluid modelling of more general two-dimensional anisotropic configurations.  The ``macroscopic'' insight allowed by this fluid description will hopefully help to better identify the characteristics of the so-called oblique Weibel-type instabilities occurring in this configuration, in which the features of Weibel, CFI and TSI modes are coupled.\\

\bibliographystyle{plain}
\bibliography{jpp_trwi}

\begin{thebibliography}{10}

\bibitem{Abramovitz}
M.~Abramovitz.
\newblock {Elementary Analytical Methods}.
\newblock In M.~Abramovitz and I.A. Stegun, editors, {\em Handbook of
  Mathematical Functions with Formulas, Graphs and Mathematical Tables}, pages
  113--138. National Bureau of Standards, 1965.

\bibitem{BasuPOP}
B.~Basu.
\newblock Moment equation description of weibel instability.
\newblock {\em {Phys. Plasmas}}, {9}({5131}), {2002}.

\bibitem{Bret05}
A.~Bret, M.~C. Firpo, and C.~Deutsch.
\newblock Electromagnetic instabilities for relativistic beam-plasma
  interaction in whole k space : nonrelativistic beam and plasma temperature
  effects.
\newblock {\em {Phys. Rev. E}}, {72}({016403}), {2005}.

\bibitem{Bret&GremilletPOPHow}
A.~Bret, L.~Gremillet, and J.C. Bellido.
\newblock How really transverse is the filamentation instability ?
\newblock {\em {Phys. Plasmas}}, {14}({032103}), {2007}.

\bibitem{Bret10}
A.~Bret, L.~Gremillet, and M.~E. Dieckmann.
\newblock Multidimensional electron beam-plasma instabilities in the
  relativistic regime.
\newblock {\em {Phys. Plasmas}}, {17}({120501}), {2010}.

\bibitem{Bret&Stockem14}
A.~Bret, A.~Stockem, R.~Narayan, and L.~O. Silva.
\newblock Collisionless weibel shocks : full formation mechanism and timing.
\newblock {\em {Phys. Plasmas}}, {21}({072301}), {2014}.

\bibitem{Califano&Attico}
F.~Califano, N.~Attico, F.~Pegoraro, G.~Bertin, and S.~V. Bulanov.
\newblock Nonlinear filamentation instability driven by an inhomogeneous
  current in a collisionless plasma.
\newblock {\em {Phys. Rev. Lett.}}, {86}({5293}), {2001}.

\bibitem{Califano&DelSartoPRL}
F.~Califano, D.~Del~Sarto, and F.~Pegoraro.
\newblock Three-dimensional magnetic structures generated by the development of
  the filamentation (weibel) instability in the relativistic regime.
\newblock {\em {Phys. Rev. Lett.}}, {96}({105008}), {2006}.

\bibitem{Califano&Pegoraro&BulanovPRE}
F.~Califano, F.~Pegoraro, and S.~V. Bulanov.
\newblock Spatial structure and time evolution of the weibel instability in
  collisionless inhomogeneous plasmas.
\newblock {\em {Phys. Rev. E}}, {56}({963}), {1997}.

\bibitem{Califano&PrandiPRE}
F.~Califano, R.~Prandi, F.~Pegoraro, and S.~V. Bulanov.
\newblock Nonlinear filamentation instability driven by an inhomogeneous
  current in a collisionless plasma.
\newblock {\em {Phys. Rev. E}}, {58}({7837}), {1998}.

\bibitem{DS_PRE}
D.~Del~Sarto, F.~Pegoraro, and F.~Califano.
\newblock Pressure anisotropy and small spatial scales induced by velocity
  shear.
\newblock {\em {Phys. Rev. E}}, {93}:{053203}, {2016}.

\bibitem{DSArxiv}
D.~Del~Sarto, F.~Pegoraro, and A.~Tenerani.
\newblock "magneto-elastic" waves in an anisotropic magnetised plasma.
\newblock {\em {preprint arXiv:1509.04938}}, {??}({??}), {2015}.

\bibitem{Franci}
L.~Franci, P.~Hellinger, L.~Matteini, A.~Verdini, and S.~Landi.
\newblock Two-dimensional hybrid simulations of kinetic plasma turbulence:
  Current and vorticity vs proton temperature.
\newblock {\em {AIP Conf. Proc.}}, {1720}({040003}), {2016}.

\bibitem{FriedPF}
B.~D. Fried.
\newblock Mechanism for instability of transverse plasma waves.
\newblock {\em {Phys. Fluids}}, {2}({337}), {1959}.

\bibitem{Fried&Conte}
B.D. Fried and S.D. Conte.
\newblock In New York~NY Academic~Press, editor, {\em The Plasma Dispersion
  Function}. 1961.

\bibitem{GhizzoTrio}
A.~Ghizzo and P.~Bertrand.
\newblock On the multi-stream approach of relativistic weibel instability. i,
  ii and iii.
\newblock {\em {Phys. Plasmas}}, {20}({082109}), {2013}.

\bibitem{GhizzoJPP}
A.~Ghizzo, M.~Sarrat, and D.~Del~Sarto.
\newblock Vlasov models for kinetic weibel-type instabilities.
\newblock {\em {to be submitted}}, {??}({??}), {in preparation}.

\bibitem{Ghorbanalilu}
M.~Ghorbanalilu, S.~Sadegzadeh, Z.~Ghaderi, and A.~R. Niknam.
\newblock Weibel instability for a streaming electron, counterstreaming e-e,
  and e-p plasmas with intrinsic temperature anisotropy.
\newblock {\em {Phys. Plasmas}}, {21}({052102}), {2014}.

\bibitem{InglebertPOP}
A.~Inglebert, A.~Ghizzo, T.~Reveill\'e, P.~Bertrand, and F.~Califano.
\newblock Electron temperature anisotropy instabilities represented by
  superposition of streams.
\newblock {\em {Phys. Plasmas}}, {19}({122109}), {2012}.

\bibitem{InglebertEPL}
A.~Inglebert, A.~Ghizzo, T.~Reveill\'e, D.~Del~Sarto, P.~Bertrand, and
  F.~Califano.
\newblock A multi-stream vlasov modeling unifying relativistic weibel-type
  instabilities.
\newblock {\em {EuroPhys. Lett.}}, {95}({45002}), {2011}.

\bibitem{LazarJPP}
M.~Lazar, M.~E. Dieckmann, and S.~Poedts.
\newblock Resonant weibel instability in counterstreaming plasmas with
  temperature anisotropies.
\newblock {\em {J.~Plasma Phys.}}, {76}({1}):{49}, {2010}.

\bibitem{LazarTAJ}
M.~Lazar, R.~Schlickeiser, R.~Wielebinski, and S.~Poedts.
\newblock Cosmological effects of weibel-type instabilities.
\newblock {\em {Astrophys. J.}}, {693}:{1133}, {2009}.

\bibitem{Masson_Laborde}
P.~E. Masson-Laborde, W.~Rozmus, Z.~Peng, D.~Pesme, S.~Hüller, M.~Casanova,
  V.~Yu. Bychenkov, T.~Chapman, and P.~Loiseau.
\newblock Evolution of the stimulated raman scattering instability in
  two-dimensional particle-in-cell simulations.
\newblock {\em {Phys. Plasmas}}, {17}({092704}), {2010}.

\bibitem{MedvedevTAJ}
M.~V. Medvedev and A.~Loeb.
\newblock Generation of magnetic fields in the relativistic shock of gamma-ray
  burst sources.
\newblock {\em {Astrophys. J.}}, {526}({697}), {1999}.

\bibitem{PegoraroScripta}
F.~Pegoraro, S.~V. Bulanov, F.~Califano, and M.~Lontano.
\newblock Nonlinear development of the weibel instability and magnetic field
  generation in collisionless plasmas.
\newblock {\em {Phys. Scripta}}, {T63}({262}), {1996}.

\bibitem{SarratEPL}
M.~Sarrat, D.~Del~Sarto, and A.~Ghizzo.
\newblock Fluid description of weibel-type instabilities via full pressure
  tensor dynamics.
\newblock {\em {EuroPhys. Lett.}}, {??}({?}), {2016}.

\bibitem{SchlickeiserTAJ}
R.~Schlickeiser and P.~K. Shukla.
\newblock Cosmological magnetic field generation by the weibel instability.
\newblock {\em {Astrophys. J.}}, {599}({57}), {2003}.

\bibitem{scudder2016collisionless}
J.~D. Scudder.
\newblock Collisionless reconnection and electron demagnetization.
\newblock In {\em Magnetic Reconnection}, pages {33--100}. {Springer}, {2016}.

\bibitem{scudder2008illuminating}
J.~D. Scudder and W.~S. Daughton.
\newblock “illuminating” electron diffusion regions of collisionless
  magnetic reconnection using electron agyrotropy.
\newblock {\em {J. Geophys. Res.-Space}}, {113}({A6}), {2008}.

\bibitem{scudder2012first}
J.~D. Scudder, R.~D. Holdaway, W.~S. Daughton, H.~Karimabadi, V.~Roytershteyn,
  C.~T. Russell, and J.~Y. Lopez.
\newblock First resolved observations of the demagnetized electron-diffusion
  region of an astrophysical magnetic-reconnection site.
\newblock {\em {Phys. Rev. Lett.}}, {108}({22}):{225005}, {2012}.

\bibitem{servidio2012local}
S~Servidio, F~Valentini, F~Califano, and P~Veltri.
\newblock Local kinetic effects in two-dimensional plasma turbulence.
\newblock {\em {Phys. Rev. Lett.}}, {108}({4}):{045001}, {2012}.

\bibitem{servidio2015kinetic}
S~Servidio, F~Valentini, D~Perrone, A~Greco, F~Califano, W.~H. Matthaeus, and
  P~Veltri.
\newblock A kinetic model of plasma turbulence.
\newblock {\em {J.~Plasma Phys.}}, {81}({01}):{325810107}, {2015}.

\bibitem{TzoufrasPRL}
M.~Tzoufras, C.~Ren, F.S. Tsung, J.W. Tonge, W.B. Mori, M.~Fiore, R.A. Fonseca,
  and L.O Silva.
\newblock Space-charge effects in the current-filamentation or weibel
  instability.
\newblock {\em {Phys. Rev. Lett.}}, {96}({105002}), {2006}.

\bibitem{WeibelPRL}
E.~S. Weibel.
\newblock Spontaneously growing transverse waves in a plasma due to an
  anisotropic velocity distribution.
\newblock {\em {Phys. Rev. Lett.}}, {2}({83}), {1959}.

\end{thebibliography}

\end{document}